\documentclass[useAMS,usenatbib]{mn2e}

\usepackage[dvips]{graphicx}
\usepackage[]{txfonts}
\usepackage{subfigure}
\usepackage{url}
\def\simless{\mathbin{\lower 3pt\hbox
{$\rlap{\raise 5pt\hbox{$\char'074$}}\mathchar"7218$}}}   
\def\simmore{\mathbin{\lower 3pt\hbox
{$\rlap{\raise 5pt\hbox{$\char'076$}}\mathchar"7218$}}}   
\newcommand{\be}{\begin{equation}}
\newcommand{\ee}{\end{equation}}
\newcommand       \bea          {\begin{eqnarray}}
\newcommand       \eea          {\end{eqnarray}}

\def\simlt{\mathrel{\hbox{\rlap{\hbox{\lower4pt\hbox{$\sim$}}}\hbox{$<$}}}}
\def\simgt{\mathrel{\hbox{\rlap{\hbox{\lower4pt\hbox{$\sim$}}}\hbox{$>$}}}}
\def\lesssim{\mathrel{\hbox{\rlap{\hbox{\lower4pt\hbox{$\sim$}}}\hbox{$<$}}}}

\def\gtrsim{\mathrel{\hbox{\rlap{\hbox{\lower4pt\hbox{$\sim$}}}\hbox{$>$}}}}

\title[Transients from Planet-Star Mergers]{Optical and X-ray Transients from Planet-Star Mergers}

\author[]{B.~D.~Metzger$^{1,3}\thanks{E-mail:
bmetzger, giannios@astro.princeton.edu, dave@ias.edu}$, D.~Giannios$^{1}$, D.~S. Spiegel$^{2}$\\
$^{1}$Department of Astrophysical Sciences, Peyton Hall, Princeton University, Princeton, NJ 08544, USA  \\ $^{2}$Institute for Advanced Study, Einstein Drive, Princeton, NJ 08540 \\ $^{3}$NASA Einstein Fellow}

\begin{document}
\date{Received / Accepted}
\pagerange{\pageref{firstpage}--\pageref{lastpage}} \pubyear{2012}

\maketitle

\label{firstpage}

\begin{abstract}

We evaluate the prompt observational signatures of the merger between a massive close-in planet (a `hot Jupiter') and its host star, events with an estimated Galactic rate of $\sim 0.1-1$ yr$^{-1}$.  Depending on the ratio of the mean density of the planet $\bar{\rho}_{\rm p}$ to that of the star $\bar{\rho}_{\star}$, a planet-star merger results in three possible outcomes.  If $\bar{\rho}_{\rm p}/\bar{\rho}_{\star} \gtrsim 5$, then the planet directly plunges below the stellar atmosphere before being disrupted by tidal forces.  The dissipation of orbital energy creates a hot wake behind the planet, producing a EUV/soft X-ray transient that increases in brightness and temperature as the planet sinks below the stellar surface.  The peak luminosity $L_{\rm EUV/X} \lesssim 10^{36}$ erg s$^{-1}$ is achieved weeks to months prior to merger, after which the stellar surface is enshrouded by an outflow driven by the merger.  The final stages of the inspiral are accompanied by an optical transient powered by the recombination of hydrogen in the outflow, which peaks at a luminosity $\sim 10^{37}-10^{38}$ erg s$^{-1}$ on a timescale $\sim$days.

If the star is instead significantly denser ($\bar{\rho}_{\rm p}/\bar{\rho}_{\star} \lesssim 5$), then the planet overflows its Roche Lobe above the stellar surface.  For $\bar{\rho}_{\rm p}/\bar{\rho}_{\star} \lesssim 1$ mass transfer is stable, resulting in the planet being accreted on the relatively slow timescale set by tidal dissipation.  However, for an intermediate density range $1 \lesssim \bar{\rho}_{\rm p}/\bar{\rho}_{\star} \lesssim 5$ mass transfer may instead be unstable, resulting in the dynamical disruption of the planet into an accretion disk around the star.  Outflows from the super-Eddington accretion disk power an optical transient with a peak luminosity $\sim 10^{37}-10^{38}$ erg s$^{-1}$ and characteristic duration $\sim$week$-$months.  Emission from the disk itself becomes visible once the accretion rate decreases below the Eddington rate, resulting in a bolometric brightening and shift of the spectral peak to UV wavelengths.  Optical transients from both direct-impact merger and tidal-disruption events in some ways resemble classical novae, but can be distinguished by their higher ejecta mass and lower velocity $\sim$hundreds km s$^{-1}$, and by hard pre- and post-cursor emission, respectively.  The most promising search strategy is with combined surveys of nearby massive galaxies (e.g.~M31) at optical, UV, and X-ray wavelengths with cadences from days to months.
    
\end{abstract} 
  

\section{Introduction} 
\label{intro}

Theoretical models of planet formation suggest that gas giants form beyond the ice line at radii $\gtrsim$ several AU from their central star (e.g.~\citealt{Inaba+03}).  It is thus a mystery how some exoplanets, colloquially known as `hot Jupiters', are transported to their current locations at radii $\lesssim 0.5$ AU (e.g.~\citealt{Ida&Lin04}).  The misalignment between the orbital planes of some hot Jupiters and the spin axes of their host stars (\citealt{Triaud+10}; \citealt{Schlaufman10}; \citealt{Winn+10}) suggests that not all migrate via interaction with the proto-stellar disk alone (although see \citealt{Foucart&Lai11}).  At least some hot Jupiters appear to have migrated via other mechanisms, such as planet-planet scattering (e.g.~\citealt{Rasio&Ford96}; \citealt{Weidenschilling&Marzari96}; \citealt{Juric&Tremaine08}) or the \citet{Kozai62} mechanism (e.g.~\citealt{Takeda&Rasio05}; \citealt{Fabrycky&Tremaine07}; \citealt{Socrates+11}).  In both cases the planet initially approaches small radii on an eccentric trajectory, before tidal dissipation circularizes the orbit.

\citet{Guillochon+11}, on the other hand, argue that the current locations of many close-in exoplanets are inconsistent with migration resulting solely due to the inward scattering from original orbits exterior to the ice line, since the small pericenter distances implied by such trajectories would have resulted either in tidal-disruption of the planet or have completely ejected it from the system.  One interpretation of their result is that some planets have migrated further {\it after} their orbit is circularized, due to tidal dissipation within the star.  

The angular momentum of most hot Jupiters is sufficiently low that no state of tidal equilibrium exists (the system is `\citealt{Darwin1880} unstable'; \citealt{Hut81}; \citealt{Levrard+09}), suggesting that the ultimate `end state' of tidal dissipation is a merger between the planet and its host star (e.g.~\citealt{Jackson+08}).  Depending on the quality factor of tidal dissipation within the star $Q'_{\star} \sim 10^{6}$, the semi-major axes of several known hot Jupiters are sufficiently small that a merger will indeed occur on a relatively short timescale $\lesssim 10-100$ Myr (e.g.~\citealt{Li+10}).  The lack of old planetary systems with very short orbital periods hints that mergers indeed result from tidal orbital decay (\citealt{Jackson+09}).  

One might expect that since $\sim 1$ per cent of stars host hot Jupiters (e.g.~\citealt{Mayor&Queloz12}), and since the Galactic star-formation rate is $\sim 1-10$ yr $^{-1}$ \citep{Naab&Ostriker06}, then the Galactic rate of planet-star mergers should be $\lesssim 0.1$ yr$^{-1}$.  However, the actual merger rate of planets with main sequence stars can in principle be significantly higher $\sim 1$ yr$^{-1}$, depending on the value of $Q'_{\star}$ and the rate at which hot Jupiters are replenished by migration from the outer stellar system (e.g.~\citealt{Socrates+11}; see $\S\ref{sec:freq}$).  A comparable or greater number of mergers may occur in evolved stellar systems, since the timescale for tidal migration $\tau_{\rm plunge} \propto R_{\star}^{-5}$ (eq.~[\ref{eq:plungetime}]) depends sensitively on stellar radius $R_{\star}$, which increases as the star evolves off the main sequence.  In fact, irrespective of the efficiency of tidal dissipation, it seems inevitable that almost all planets with main-sequence orbital separation $\lesssim 1$ AU eventually merger with their stars, since stars expand dramatically during the red giant and asymptotic giant branch phases of evolution (although post-main-sequence mergers are significantly less energetic events than their main sequence brethren).   
 
In this paper we explore the direct observational consequences of the merger of gas giant planets with their host stars.  Depending on the properties of the planet-star system, we show that the merger event can be violent, resulting a bright and long-lived electromagnetic signature.  Given the rapid pace of recent advances in our knowledge of exoplanets, and the advent of sensitive wide-field surveys across the electromagnetic spectrum, now is an optimal time to evaluate the transient signatures of planet-star mergers and how best to go about detecting them.

This paper is organized as follows.  In $\S\ref{sec:freq}$ we summarize the properties of known hot Jupiters and use them to estimate the rate of planet-star mergers in galaxies similar to the Milky Way.  In \S\ref{sec:outcomes} we identify three qualitatively different merger outcomes, which are distinguished in part by whether the planet fills its Roche radius at an orbital separation $a_{\rm t}$ which is larger or smaller, respectively, than the stellar radius $R_{\star}$.  If $a_{t} \lesssim R_{\star}$ then the planet is not disrupted until it has fallen below the stellar photosphere.  In $\S\ref{sec:interior}$ we describe the dynamics of the planetary inspiral and argue that such `direct-impact' merger events are accompanied by a extreme ultraviolet (EUV)/soft X-ray transient that originates from the hot gas created behind the planet as it penetrates the stellar atmosphere ($\S\ref{sec:Xray}$).  This emission, which ceases weeks$-$months prior to merger, is immediately followed by rising optical emission powered by the recombination of hydrogen in an outflow from the stellar surface, which peaks on a timescale $\sim$days after the planet plunges below the stellar surface ($\S\ref{sec:optical}$). 

If $a_{t} \gtrsim R_{\star}$, then the planet instead overflows its Roche radius above the stellar surface.  If mass transfer is stable, then the planet is consumed on the relatively long timescale set by the rate of tidal dissipation; this results in a relatively short-lived mass-transferring binary, but is unlikely to produce a bright electromagnetic display.  If, on the other hand, mass transfer is unstable (or if the initial planet trajectory is eccentric), then the planet is disrupted into an accretion disk just outside the stellar surface (`tidal-disruption event').  Thermal radiation from the disk, and from super-Eddington outflows from its surface, powers a bright optical-UV transient ($\S\ref{sec:exterior}$).  In $\S\ref{sec:discussion}$ we discuss our results, including a concise summary of the predicted transients from planet-star mergers ($\S\ref{sec:summary}$); detection prospects and strategies, in particular with surveys of nearby galaxies at optical, UV, and X-ray wavelengths ($\S\ref{sec:detect}$); and a brief conclusion, which focuses on unresolved theoretical issues and future work ($\S\ref{sec:conclusions}$). 

Our work here builds on, and is broadly consistent with, past theoretical work into the observational signatures of binary stellar mergers (e.g.~\citealt{Tylenda&Soker06}; \citealt{Soker&Tylenda06}), events which \citet{Soker&Tylenda06} refer to as `mergebursts' (see also \citealt{Bear+11}).

\section{Galactic rate of Planet-Star Mergers}
\label{sec:freq}

Dissipation of tides raised on a planet and on its host star can act to change their orbital separation.  Two important categories of tidal interactions can be termed ``circularization tides'' and ``synchronization tides.''  Dissipation of the former kind acts to damp eccentricity at constant orbital angular momentum $\mathcal{L}$.  Since $\mathcal{L} \propto \sqrt{a(1-e^2)}$, where $a$ is semi-major axis and $e$ is eccentricity, circularization of the orbit ($e \rightarrow 0$) at constant $\mathcal{L}$ shrinks the binary separation.  

Dissipation of synchronization tides instead occurs due to asynchronous rotation of the bodies with respect to the orbital mean motion, even in the absence of eccentricity.  In particular, if a planet orbits faster than its host star rotates, then tidal torques transfer angular momentum from the planet to the star, leading to orbital decay (\citealt{Darwin1880}; \citealt{Hut81}; \citealt{Soker&Tylenda06}).  The characteristic timescale for this process (the tidal `plunging time' $\tau_{\rm plunge}$) is found by integrating the tidal dissipation equations assuming a slowly rotating star (\citealt{Levrard+09}; \citealt{Brown+11}), which can be estimated as (see also Appendix \ref{app:B})
\begin{eqnarray}
\label{eq:plungetime} 
\tau_{\rm plunge} & \sim & \frac{4}{117} \frac{Q'_\star}{n} \left( \frac{a}{R_\star} \right)^5 \frac{M_\star}{M_{\rm p}} \\
\nonumber & \approx & 1.5\times 10^{5}{\rm\,yr}\left(\frac{Q'_\star}{10^6} \right)\left(\frac{M_{\star}}{M_{\odot}}\right)^{1/2}\left(\frac{a}{2R_{\odot}}\right)^{13/2} \left( \frac{R_\star}{R_\odot} \right)^{-5}\left( \frac{M_{\rm p}}{M_{\rm J}} \right)^{-1},
\end{eqnarray}
where $Q'_{\star} \equiv 3Q_{\star}/k_2$ is the modified tidal quality factor of the star ($Q_{\star}$ is the specific tidal dissipation function and $k_2$ is the Love number); $n \equiv 2\pi / P \simeq (GM_{\star}/a^{3})^{1/2}$ is the orbital mean motion; $P$ is the orbital period; $a$ is the orbital semi-major axis; $R_{\star}$ is the stellar radius; and $M_\star$ and $M_{\rm p}$ are the stellar and planetary masses, respectively, where the latter is normalized to the mass of Jupiter $M_{\rm J}$.

\begin{figure}
\resizebox{\hsize}{!}{\includegraphics[angle=0]{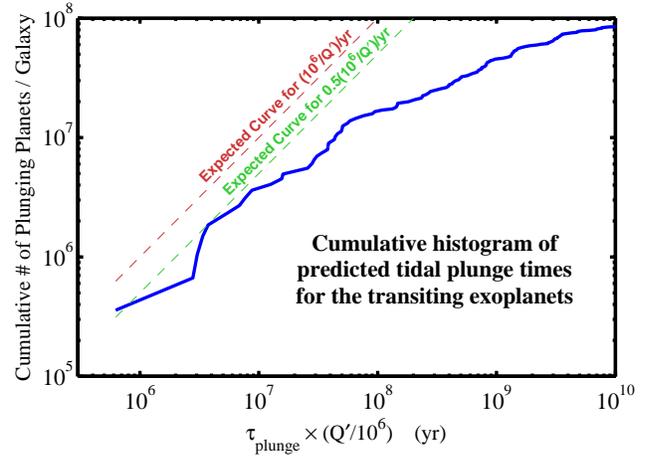}}
\caption[] {Cumulative number of planet-star mergers in the Galaxy ({\it solid blue line}) as a function of plunging time $\tau_{\rm plunge}$ (eq.~[\ref{eq:plungetime}]), the latter normalized to a fiducial value $Q'_{\star} \sim 10^{6}$ for the rate of tidal dissipation in the star.  Overlaid dashed lines indicate the expected distributions for steady-state merger rates of $(10^6/Q'_{\star})$ yr$^{-1}$ ({\it red}) and $0.5(10^6/Q'_{\star})$ yr$^{-1}$ ({\it green}), respectively.  Deviations of the model predictions from the observed distribution at larger values of $\tau_{\rm plunge}$ indicate the presence of additional effects not included in our model, such as a dependence of $Q'_{\star}$ on orbital period or a break-down of the steady state assumption (see text for further details).}
\label{fig:cumsum}
\end{figure}

Figure~\ref{fig:cumsum} shows the cumulative number of planets as a function of tidal plunge time, calculated by applying equation (\ref{eq:plungetime}) to the distribution of $\sim 160$ known transiting extrasolar planets (\citealt{Schneider+11}).\footnote{\url{http://exoplanet.eu}}  This $\tau_{\rm plunge}$ distribution can be used to estimate the rate of planet-star mergers if one can extrapolate the known sample of exoplanets to those of the galaxy as a whole.  In order to make this conversion, we assume that the observational selection function is given simply by the geometric probability of transit ($R_\star/a$).  We also assume that a fraction $\sim 10^{-5}$ of all stars in the Milky Way have been searched by transit surveys (G. Bakos, private communication), such that each transiting planet used in Figure \ref{fig:cumsum} ``counts'' as $10^5(a/R_{\star})$ planets in the Galaxy.  

Dashed lines in Figure \ref{fig:cumsum} represent the expected plunge-time distribution if one assumes that mergers occur with a steady-state rate of $10^6/Q'_{\star}$ yr$^{-1}$ and $0.5(10^6/Q'_{\star})$ yr$^{-1}$, respectively.  A comparison of the observed distribution to these models indicate that, at low values of plunge time $\sim 10^{6-7}(Q'_{\star}/10^6)$~yrs, roughly one merger occurs in the Galaxy every $2(Q'_{\star}/10^6)$ years.  Obvious deviations of the cumulative rate from the model predictions occur at large values $\tau_{\rm plunge}$.  These indicate the onset of additional effects not included in our simple model, such as a dependence of $Q'_{\star}$ on orbital period (e.g.~\citealt{Goodman&Dickson98}); the onset of additional selection effects; or a break-down of the steady-state assumption.  The latter could result from the ongoing injection of planets from larger radii (e.g.~\citealt{Socrates+11}), or by insitu planet formation following a binary star merger (e.g.~\citealt{Martin+11}).  Although our estimate of the merger rate from Figure \ref{fig:cumsum} is uncertain by at least an order of magnitude, the fiducial range of values $\sim 0.1-1$ yr$^{-1}$ that we find are still sufficiently high that transient surveys with duration comparable to this mean interval $\sim$year---decade could in principle detect such events within the Milky Way or nearby galaxies, provided that the resulting emission is sufficiently bright.  


As mentioned in the Introduction, the rate of planet-star mergers could be enhanced as a result of post main sequence evolution.  After the star evolves off of the main sequence, its larger radius $R_\star$ significantly decreases the plunging timescale $\tau_{\rm plunge} \propto R_{\star}^{-5}$ (eq.~[\ref{eq:plungetime}]; even in the subgiant phase, for which $R_{\star} \sim $ few $R_{\odot}$), more than compensating for the shorter duration of post main sequence evolution.  This suggests that a sizable fraction of all stars with semi-major axes within $a \lesssim $ 0.1AU may be consumed (\citealt{Sato+08}; \citealt{Hansen10}; \citealt{Lloyd11}).  Since the fraction of stars with planets in this radius range approaches unity \citep{Mayor+11}, the Galactic rate of planet-star mergers may approach a significant fraction of the star-formation rate $\sim 1-10$ yr$^{-1}$.  

Mergers are likely to be even more common during the giant phase when the stellar radius expands to $R_{\star} \sim$AU (e.g.~\citealt{Siess&Livio99a,Siess&Livio99b}; \citealt{Retter&Marom03}).  The low density of the stellar envelope in this case would almost certainly result in what we define as a direct-impact merger ($\S\ref{sec:interior}$).  However, the much lower gravitational energy released near the stellar surface in such an event as compared to the merger with a dwarf or subgiant results in much fainter transient emission that is more challenging to detect (although these events may have other observable consequences; \citealt{Retter&Marom03}; \citealt{Nordhaus+10}; \citealt{Nordhaus+11}).  For this reason we focus on planet mergers with compact stars, even though their total rate may be somewhat lower.

\section{Three Possible Merger Outcomes}
\label{sec:outcomes}

\begin{figure}
\resizebox{\hsize}{!}{\includegraphics[angle=0]{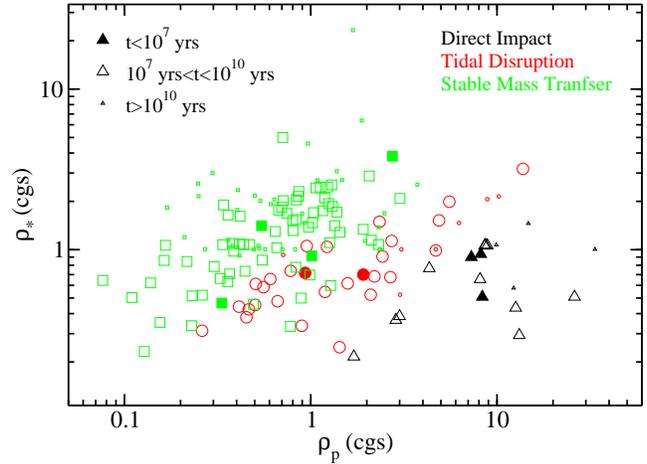}}
\caption[] {Sample of 160 transiting extra-solar planets in the space of mean stellar density $\bar{\rho}_{\star}$ versus mean planetary density $\bar{\rho}_{\rm p}$.  The plunge time of each planet $\tau_{\rm plunge}$ (eq.~[\ref{eq:plungetime}]) is denoted by the size and fill of each marker.  The shape of each marker indicates the predicted fate of each system as described in $\S\ref{sec:outcomes}$ and Appendix \ref{app:A}, assuming that the mass and radii of the planet and star are the same upon merger as their current values.  Black triangles indicate planets that directly impact the stellar surface ($\bar{\rho}_{\rm p}/\bar{\rho}_{\star} \gtrsim 5$; `direct-impact' merger events; $\S\ref{sec:interior}$); green squares to planets that undergo stable mass transfer above the surface ($\bar{\rho}_{\rm p}/\bar{\rho}_{\star} \lesssim 1$); and red circles to planets that undergo unstable mass transfer and are shredded into an accretion disk around the star ($1 \lesssim \bar{\rho}_{\rm p}/\bar{\rho}_{\star} \lesssim 5$; `tidal-disruption events'; $\S\ref{sec:exterior}$).}
\label{fig:demographics}
\end{figure}

So long as no mass transfer occurs between the planet and the star, tidal forces control the rate at which the orbit decays.  As the planet inspirals towards the star, its fate depends on whether it fills its Roche-lobe above or below the stellar surface.  In the latter case the planet plunges directly into the stellar atmosphere.  This produces an event that we define as a `direct-impact' merger event, the result of which we describe in detail in $\S\ref{sec:interior}$.  If Roche-lobe overflow instead begins above the stellar surface, then the planet is either dynamically disrupted into a compact disk around the star (`tidal-disruption' merger event; $\S\ref{sec:exterior}$), or it gradually transfers its mass over a much longer timescale (stable mass transfer).     

The volume of the Roche-lobe of the planet in a synchronous orbit about a star of much greater mass is approximately given by $V_{\rm RL}\simeq 0.5 (M_{\rm p}/M_{\star})a^3$ \citep{Dai&Blandford11}.  Roche-lobe overflow thus begins when the orbital semi-major axis reaches the value
\be
a_t\simeq 2R_{\star}(\bar{\rho}_{\star}/\bar{\rho}_{\rm p})^{1/3},
\label{eq:at}
\ee
where $\bar{\rho}_{\star} = 3M_{\star}/4\pi R_{\star}^{3}$ and $\bar{\rho}_{\rm p} = 3M_{\rm p}/4\pi R_{\rm p}^{3}$ are the mean density of the star and planet, respectively. 

At the orbital separation given by equation (\ref{eq:at}) the L1 Lagrange point is located at a radial distance $X_t \simeq 0.7 (M_{\rm p}/M_{\star})^{1/3}a_t\simeq 1.4 R_{\rm p}$ from the center of mass of the planet.  The condition that Roche-lobe overflow occurs entirely below the stellar surface is thus given by $a_t<R_{\star}+X_t$, or\be
2 (\bar{\rho}_{\star}/\bar{\rho}_{\rm p})^{1/3}<1+1.4(R_{\rm p}/R_{\star}).
\label{eq:cond1}
\ee 
If equation (\ref{eq:cond1}) is satisfied, then a direct-impact merger occurs ($\S\ref{sec:interior}$).

If Roche-lobe overflow instead occurs outside the star (i.e.~if eq.~[\ref{eq:cond1}] is {\it not} satisfied), then the evolution of the system is more subtle.  In this case mass transfer is stable(unstable) when mass loss from the planet results in the Roche-lobe growing faster(slower) than the radius of the planet.  Here we describe the qualitative distinction between the system evolution in these two cases, the details of which are provided in Appendix \ref{app:A}.

\begin{table}
\begin{center}
\vspace{0.05 in}\caption{Outcomes of planet-star mergers based on known transiting planets}
\label{table:outcomes}
\resizebox{8cm}{!}
{
\begin{tabular}{ccc}
\hline
\hline
\multicolumn{1}{c}{Planet Sample} &
\multicolumn{1}{c}{Merger Outcome} &
\multicolumn{1}{c}{Fraction ($\%$)} \\
\hline
$\tau_{\rm plunge} < (Q'_{\star}/10^{6})10^{7}$ yrs & Direct-Impact & 33 \\
- & Tidal-Disruption & 22 \\
- & Stable Mass Transfer & 44 \\
\hline
$\tau_{\rm plunge} < (Q'_{\star}/10^{6})10^{10}$ yrs & Direct-Impact & 13 \\
- & Tidal-Disruption & 25 \\
- & Stable Mass Transfer & 62 \\
\hline
All Systems & Direct-Impact & 12 \\
- & Tidal-Disruption & 22 \\
- & Stable Mass Transfer & 66 \\
\hline
\hline
\end{tabular}
}
\end{center}
\end{table}

If overflow occurs when the orbital separation is sufficiently large, then material crossing L1 has sufficient specific angular momentum to form a disk well above the stellar surface.  Viscous redistribution of angular momentum within the disk causes most of its mass to accrete onto the star, whereas the majority of the angular momentum is transported to large radii where it can be transferred back into the orbit via tides exerted by the planet.  Since mass is transferred from the less massive planet to the more massive star, this (approximate) conservation of orbital angular momentum acts to widen the orbital separation.  Since the Roche-lobe volume $V_{\rm RL}\propto a^3$ increases rapidly as the orbit expands, mass transfer is temporarily halted.  Orbital decay due to tidal dissipation then again drives the system together, resulting in a series of overflow episodes that slowly exhaust the planetary material on the tidal decay timescale $\tau_{\rm plunge}(a = a_{t} \gtrsim R_{\star}) \gtrsim 10^{3}$ yr (eq.~[\ref{eq:plungetime}]).  This is the stable mass transfer case.

If, on the other hand, Roche-lobe overflow occurs only at a sufficiently small planet-star separation, then material crossing L1 plunges directly onto, or circularizes just above, the stellar surface.  In this case the bulk of the mass and angular momentum of the accreting material is deposited into the stellar envelope, which is capable of accepting it due to its considerable inertia.  Since orbital angular momentum is lost to the star, the semi-major axis of the orbit (and hence the Roche volume of the planet) remains unchanged, whereas the planet expands upon mass loss (Appendix \ref{app:A}).  This results in unstable mass transfer and hence to dynamical tidal-disruption of the planet on a characteristic timescale of several orbital periods $\sim$hours ($\S\ref{sec:exterior}$).    

Figure \ref{fig:demographics} shows the sample of transiting planets used in Figure \ref{fig:cumsum} in the space of mean stellar density $\bar{\rho}_{\star}$ versus mean planetary density $\bar{\rho}_{\rm p}$.  Each planet is marked according to both the value of its tidal plunging time (eq.~[\ref{eq:plungetime}]) and by our best estimate of its ultimate fate upon merger (`direct-impact'; `stable mass transfer'; or `tidal-disruption').  The criteria for the latter are evaluated assuming that the current masses and radii of the planets and their host stars are identical to those at the time of merger (the validity of which we discuss in $\S\ref{sec:conclusions}$).  We find that if the density ratio is sufficiently high ($\bar{\rho}_{\rm p}/\bar{\rho}_{\star}\simmore 5$), then the the planet is disrupted below the stellar surface (direct-impact mergers).  Planets somewhat denser than their host star ($1\simless\bar{\rho}_{\rm p}/\bar{\rho}_{\star}\simless 5$) will instead be dynamically disrupted close to (but outside) the stellar surface (tidal-disruption).  Finally, the least dense planets ($\bar{\rho}_{\rm p}/\bar{\rho}_{\star}\lesssim 1$) typically undergo stable mass transfer.  

Table \ref{table:outcomes} summarizes the percentage of known transiting planetary systems that are predicted to experience various outcomes upon merger.  Out of the entire sample of planets in Figure \ref{fig:demographics}, we estimate that $\sim 12$ per cent undergo a direct-impact merger.  This fraction, however, rises to $\sim 33$ per cent (3 out of 9) when considering just those systems with the shortest lifetimes $\tau_{\rm plunge} <(Q'_{\star}/10^{6})10^7$ yrs.  We estimate that $\sim 1/4$ of systems will undergo unstable mass transfer and tidal-disruption, while the remaining (approximately half to two thirds) are consumed less violently via stable mass transfer.  Note again that our estimates are implicitly restricted to the merger of planets with main sequence stars.  If one also includes mergers during the post main sequence, then the expansion of the star decreases the value of $\rho_{\star}$ and hence would increase the fraction of planets that undergo direct-impact or tidal-disruption. 

\section{Direct-Impact Merger}
\label{sec:interior}

If the planet is disrupted well inside the star ($\bar{\rho}_{\rm p} \gtrsim 5\bar{\rho}_{\star}$), then most of the gravitational energy released during this process, $E_{\rm grav} \gtrsim GM_{\star}M_{\rm p}/R_{\star} =  4 \times 10^{46} (M_{\rm p}/M_{\rm J}) $ ergs, goes into heating the stellar interior.  This additional energy will most likely act to enhance the stellar luminosity (however, see \citealt{Podsiadlowski03}), but only on the cooling timescale of the envelope at the disruption radius.  In the Sun, for example, the radiative cooling timescale at $r \sim 0.5R_{\odot}$ (where the mean interior density is $\bar{\rho} \sim 10$ g cm$^{-3}$) is\footnote{This may be an overestimate of the true cooling timescale, since heating from the disruption will increase the entropy near the point of disruption, thereby altering the stellar structure.} $t_{\rm cool} \sim 10^{6}$ yr.  Disruption of a Jupiter mass planet would thus increase the stellar luminosity by an amount $L \sim E_{\rm grav}/t_{\rm cool} \gtrsim 10^{33}$ erg s$^{-1}$, comparable to the normal solar luminosity.  Thus one characteristic of direct-impact mergers is a long-term change in the stellar evolution and appearance (\citealt{Podsiadlowski03}; \citealt{Tylenda05}).  Unfortunately, the amplitude of this variation is typically modest and its duration is typically longer than is accessible to transient surveys.  

In this section we instead describe the more prompt signature that originates as the planet first interacts with the external layers of the star, before disappearing into the interior.  The planet impacts the stellar surface on a grazing orbit with the Keplerian velocity $v_{\rm k} = (GM_{\star}/R_{\star})^{1/2}$.  This leaves a hot trail of shocked stellar gas behind it with a characteristic virial temperature $T_{\rm vir} \sim 10^{6}-10^7$ K, resulting in a powerful UV/X-ray transient for a duration of weeks$-$months (we show below that the spectrum peaks at a somewhat lower temperature than $T_{\rm vir}$).  In what follows we assume that the orbit of the planet is nearly circular at the time of merger, which, as we show in Appendix \ref{app:B}, is justified for the sample of known transiting extrasolar planets.  However, in $\S\ref{sec:discussion}$ we briefly describe how our conclusions change if the planet instead impacts the star on a highly eccentric orbit.  

\subsection{Inspiral dynamics}
\label{sec:dynamics}

\begin{figure*}
\resizebox{17cm}{!}{\includegraphics[angle=0]{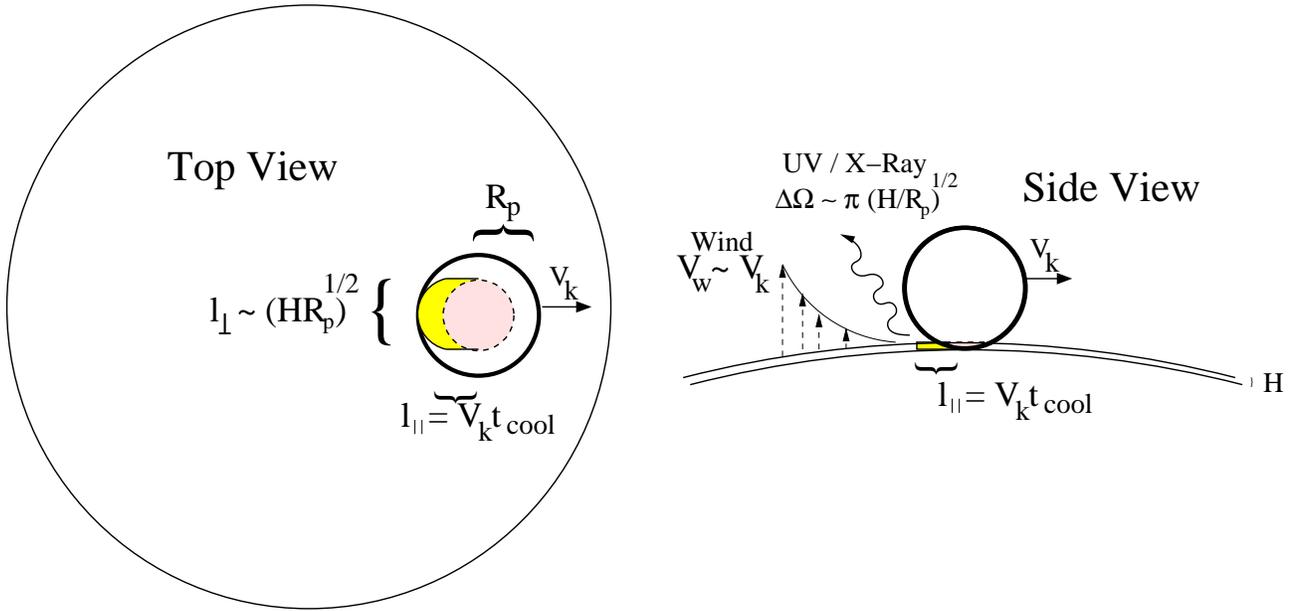}}
\caption[] {Schematic diagram of the direct-impact of a planet with the stellar surface ($\S\ref{sec:interior}$).  Show in top and side views is the geometry of the hot layer left behind as the planet grazes the stellar atmosphere ({\it yellow}).  The latitudinal width of the wake perpendicular to the planet $l_{\perp}$ is set by the width of the planet at the depth of the stellar scale height $H$.  The length of the wake parallel to the planetary motion is set by the distance $l_{\parallel} = v_{\rm k}t_{\rm cool}$ that the planet passes in a cooling time $t_{\rm cool}$, with a typical distance of several scale heights.  An (uncertain) fraction $\epsilon_{\rm rad} < 1$ of the thermal power $\dot{E}$ generated as the planet sinks through the atmosphere (eq.~[\ref{eq:edot}]) is radiated directly from the surface (typically EUV/soft X-rays; see eq.~[\ref{eq:Teff}]), while the rest of the energy (likely the majority) goes into driving an outflow further downstream (see Fig.~\ref{fig:schematicwind} for a schematic diagram of the outflow geometry).}
\label{fig:schematic}
\end{figure*}

In this section we describe the dynamics of the inspiral of the planet through the stellar atmosphere.  Then below we address the properties of the heated stellar surface ($\S\ref{sec:hotlayer}$) and the resulting electromagnetic emission from the merger ($\S\ref{sec:radiation}$).  

As the planet first encounters the stellar atmosphere of density $\rho_{\star}$, it loses angular momentum and energy via turbulent drag, causing the planet to slowly spiral inwards with time.  The drag force per unit area is given by $f_{\rm d} = \mathcal{C}_{\rm d}\rho_{\star}v_{\rm k}^{2}/2$, where $\mathcal{C}_{\rm d}$ is the dimensionless drag coefficient, which is of order unity for a supersonic flow with a high Reynolds number.  The atmospheric density decreases rapidly with radius, with a (unperturbed) scale height $H \sim 10^{-3}-10^{-2}R_{\star}$ in the outer layers which is much less than the planetary radius $R_{\rm p} \approx 0.1R_{\star}$.  The torque on the planet is thus dominated by the fluid that intercepts the planet within a distance $\approx H$ from its inner edge closest to the star.  Simple geometry\footnote{We neglect gravitational focusing, since this only increases the effective cross section of the planet by a factor $\sim M_{\rm p}R_{\star}/M_{\star}R_{\rm p} \ll 1$.} shows that in the limit that $H \ll R_{\rm p}$, the effective cross section of the planet for angular momentum loss is approximately given by $A_{\rm p} \approx H\times L_{\perp}$, where $L_{\perp} \simeq R_{\rm p}^{1/2}H^{1/2}$ is the characteristic width of the star perpendicular to the direction of motion at the penetration depth $H$ (see Fig.~\ref{fig:schematic}).  This expression is accurate for $H \lesssim R_{\rm p}$, as is valid through the final stages of the inspiral.  

In calculating the torque on the planet, we are justified in using an unperturbed value for the atmospheric scale height because the cooling timescale of the hot gas behind the planet is typically shorter than the orbital period (see below), such that the stellar atmosphere should have approximately returned to its original state by the time that the planet has passed around the star.  We are also justified in neglecting the effects of the gravitational field of the planet on the stellar scale height since the gravity of the star typically dominates near the stellar surface.  Even if the gravity of the planet is important, the planet-crossing timescale $\sim R_{\rm p}/v_{\rm k} \sim 10^{2}$ s is generally shorter than the timescale required for the atmosphere to adjust thermally $\sim H/c_{\rm s} \sim 10^{3}(H/10^{-2}R_{\odot})$ s.  Likewise, we neglect the effects of nonaxisymmetric gravitational tides raised by the passing planet (\citealt{Ricker&Taam08,Ricker&Taam12}), assuming that they are negligible compared to other forms of drag.

The drag torque $f_{\rm d}A_{\rm p}\times R_{\star}$ extracts angular momentum from the planet at the rate $\dot{J} \simeq M_{\rm p}v_{\rm k}v_{r}/2$, resulting in an inward velocity given by
\be
v_{r} \approx \mathcal{C}_{\rm d}\rho_{\star}R_{\rm p}^{1/2}H^{3/2}R_{\star}M_{\rm p}^{-1}v_{\rm k}.
\label{eq:vr}
\ee
One may define the characteristic `infall' time $t_{\rm in,drag}$ from any radius as that required for the planet to migrate inwards by a scale height, since this is the distance to encounter denser gas (and hence substantially stronger drag):
\begin{eqnarray}
t_{\rm in,drag} &\approx& 
\frac{H}{v_{r}} \approx \mathcal{C}_{\rm d}^{-1}\frac{M_{\rm p}}{\rho_{\star}R_{\star}^{2}H^{1/2}R_{\rm p}^{1/2}}\frac{R_{\star}}{v_{\rm k}} \nonumber \\
&\approx& 30\left(\frac{\rho_{\star}}{10^{-3}\rm g\,cm^{-3}}\right)^{-1}\left(\frac{M_{\rm p}}{M_{\rm J}}\right)\left(\frac{H}{10^{-2}R_{\odot}}\right)^{-1/2}t_{\rm orb},
\label{eq:tin}
\end{eqnarray}
where $t_{\rm orb} = 2\pi R_{\star}/v_{\rm k} \approx 10^{4}$ s is the orbital time at the stellar surface and in the second expression (and hereafter) we assume that $R_{\star} = R_{\odot}$, $R_{\rm p} = 0.1R_{\odot}$, and $\mathcal{C}_{\rm d} = 1$.  Equation (\ref{eq:tin}) shows that the infall time is much longer than an orbital time in the outer layers of the star where the stellar density is low, but that the rate of infall increases with time as the planet plunges deeper, such that $t_{\rm in,drag} \lesssim t_{\rm orb}$ (dynamical plunge) for $\rho_{\star} \gtrsim 0.01-0.1$ g cm$^{-3}$.  Infall is clearly an exponential runaway process.

The stellar density above which inward migration due to gas drag dominates over that due to tidal dissipation can be estimated by comparing $t_{\rm in,drag}$ to the corresponding infall time due to tidal migration
\begin{eqnarray}
t_{\rm in,tidal}\equiv \left(\frac{H}{R_{\star}}\right)\tau_{\rm plunge}|_{a = R_{\star}} \approx 17{\rm\,yr\,}\left(\frac{H}{10^{-2}R_{\odot}}\right)\left(\frac{Q'_{\star}}{10^{6}}\right)\left(\frac{M_{\rm p}}{M_{\rm J}}\right)^{-1},
\label{eq:tintidal}
\end{eqnarray}
where the factor $(H/R_{\star})$ corrects for the fact that $\tau_{\rm plunge}$ is defined in equation (\ref{eq:plungetime}) as the migration time across a distance of order the orbital separation $a \approx R_{\star}$.  By equating (\ref{eq:tin}) and (\ref{eq:tintidal}) using a realistic profile $H(\rho_{\star})$ corresponding to a main sequence solar type star (see below), we find that gas drag dominates over tidal drag (assuming $Q'_{\star} = 10^{6}$) at stellar densities greater than $\rho_{\star} \gtrsim 10^{-5}(10^{-3.5})$ g cm$^{-3}$ in the case of a planet of mass $M_{\rm p} = 1(10)M_{\odot}$.  In both cases this first occurs at a characteristic infall time of $t_{\rm in,drag} \approx t_{\rm in,tidal} \sim 100$ days.
 
The rate at which energy is dissipated by the inward migration of the planet is given by
\begin{eqnarray}
\dot{E} &\simeq& \frac{GM_{\star}M_{\rm p}v_{r}}{2R_{\star}^{2}} \nonumber \\
&\approx& 6\times 10^{37}{\rm erg\,s^{-1}}\left(\frac{\rho_{\star}}{10^{-3}\rm g\,cm^{-3}}\right)\left(\frac{H}{10^{-2}R_{\odot}}\right)^{3/2} \nonumber \\
&\approx& 2\times 10^{39}{\rm erg\,s^{-1}}\left(\frac{H}{10^{-2}R_{\odot}}\right)\left(\frac{M_{\rm p}}{M_{\rm J}}\right)\left(\frac{t_{\rm in}}{t_{\rm orb}}\right)^{-1},
\label{eq:edot}
\end{eqnarray}  
where we have used equation (\ref{eq:vr}) for $v_{r}$.  In the last expression we have substituted the infall time $t_{\rm in,drag}$ from equation (\ref{eq:tin}), so this expression is valid only when $t_{\rm in,drag} < t_{\rm in,tidal}$.

Figure \ref{fig:dynamics} shows several quantities relevant to the dynamics of the inspiral of a hot Jupiter planet into a solar mass star, as a function of the increasing stellar density $\rho_{\star}$ encountered as the planet moves inwards.  These include the density scale height of the stellar atmosphere $H$; the dissipated power $\dot{E}$ (eq.~[\ref{eq:edot}]); and the planet infall time $t_{\rm in} \equiv $ min[$t_{\rm in,tidal},t_{\rm in,drag}$] (eqs.~[\ref{eq:tin},\ref{eq:tintidal}]) assuming $Q'_{\star} = 10^{6}$, normalized to the orbital time $t_{\rm orb} \approx 10^{4}$ s.  For $t_{\rm in}$ two cases are shown, corresponding to different planet masses $M_{\rm p} = 1M_{\rm J}$ and $10M_{\rm J}$, respectively.  The times at which the inward migration rate due to gas drag exceeds that due to tidal dissipation (i.e.~$t_{\rm in,drag}< t_{\rm in,tidal}$) are marked in the top panel with vertical dotted red lines.  Our calculations employ the stellar structure $H(\rho_{\star})$ of a 1$M_{\odot}$ ZAMS star at a characteristic age $t = 5$ Gyr, as calculated using the EZ stellar evolution code \citep{Paxton04}. 

Figure \ref{fig:dynamics} shows that as the stellar density increases from $\sim10^{-5}(10^{-4})$g cm$^{-3}$ to $\sim10^{-2}(10^{-1})$ g cm$^{-3}$, the infall time $t_{\rm in}$ decreases from $\sim10^{3}t_{\rm orb} \sim$100 days to dynamical infall $\sim t_{\rm orb} \sim$hrs, for planetary masses $M_{\rm p} = 1M_{\rm J} (10M_{\rm J})$, respectively.  The dissipated power increases over this timescale from $\sim 10^{33}$ erg s$^{-1}$ to $\sim 10^{40}$ erg s$^{-1}$ in the $M_{\rm p} = 1M_{\rm J}$ case.  The evolution in the $M_{\rm p} = 10M_{\rm J}$ case is similar, except that the final power at the point of dynamical infall reaches a higher value $\sim 10^{41}$ erg s$^{-1}$ due to the greater gravitational energy released (eq.~[\ref{eq:edot}]).  Although the rate of dissipated power becomes very high during the final stages of the merger, below we show that $\dot{E}$ represents only an upper limit on the radiated electromagnetic power.

\subsection{Properties of the Hot Post-Planetary Layer}
\label{sec:hotlayer}

\begin{figure*}
\resizebox{\hsize}{!}{\includegraphics[angle=0]{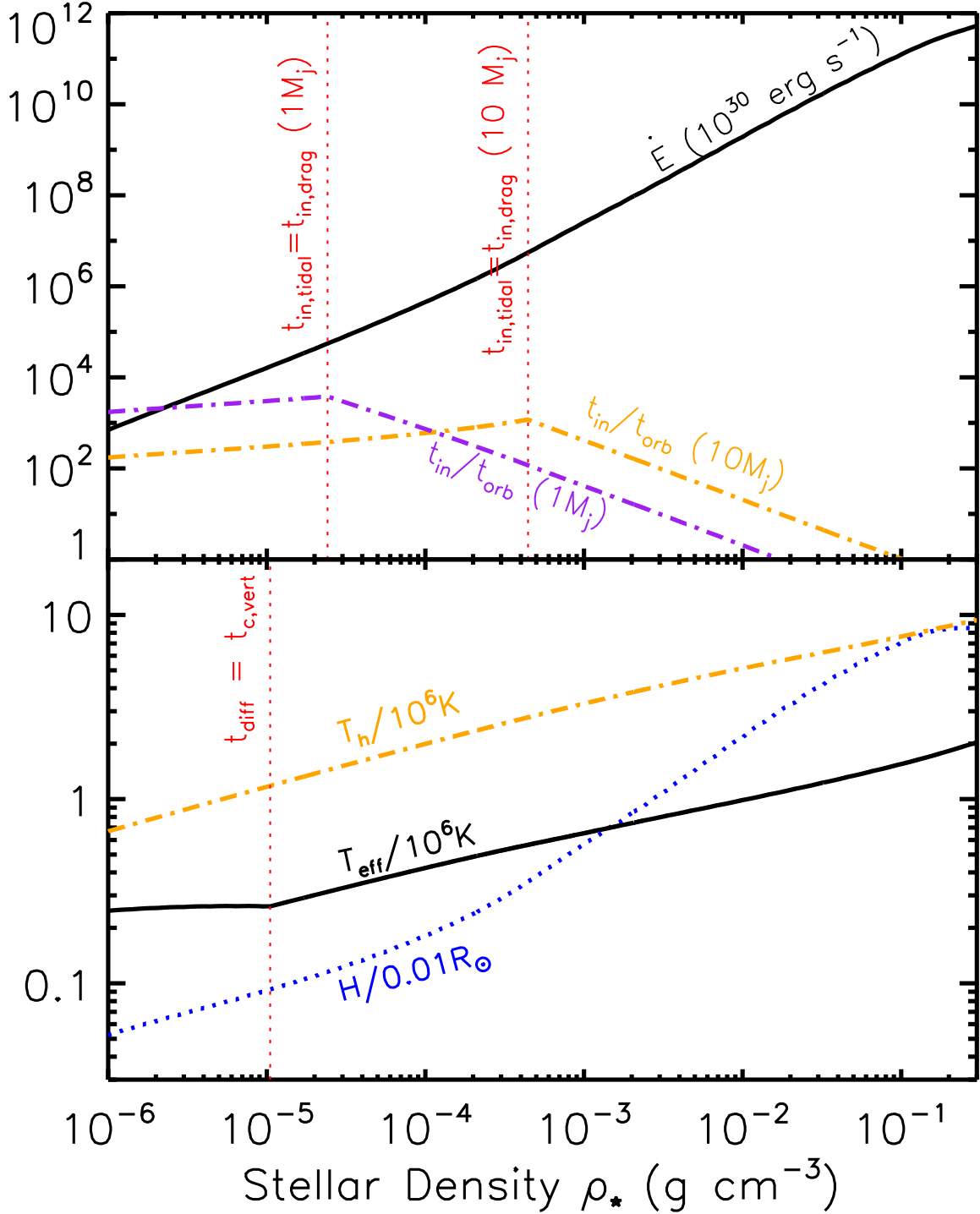}}
\caption[] {Quantities relevant to the dynamics of the inspiral of a hot Jupiter planet into a solar mass star in the case of a direct-impact merger ($\S\ref{sec:interior}$), as a function of the increasing stellar density $\rho_{\star}$ encountered as the planet moves inwards.  Quantities shown include the stellar density scale height $H$ ({\it dotted blue line, bottom panel}); the dissipated power $\dot{E}$ (eq.~[\ref{eq:edot}]; {\it dark solid line, top panel}); and the planet infall time $t_{\rm in}$, in units of the orbital time $t_{\rm orb} \approx 10^{4}$ s (eqs.~[\ref{eq:tin},\ref{eq:tintidal}]; {\it dot-dashed line, top panel}).  For $t_{\rm in}$ two cases are shown, corresponding to different planet masses $M_{\rm p} = 1M_{\rm J}$ ({\it purple}) and $10M_{\rm J}$ ({\it orange}), respectively.  Times after which the inward migration timescale due to gas drag $t_{\rm in, drag}$ (eq.~[\ref{eq:tin}]) is less than that due to tidal dissipation $t_{\rm in,tidal}$ (eq.~[\ref{eq:tintidal}], assuming $Q'_{\star} = 10^{6}$) are shown in the top panel with vertical dotted red lines.  Also shown on the bottom panel are the temperature $T_{\rm h}$ of the hot layer on the stellar surface behind the passing planet ({\it dark solid line}) and the effective temperature $T_{\rm eff}$ of radiation from the hot layer ({\it orange dot-dashed line}).  The stellar structure $H(\rho_{\star})$ used in our calculation is that of a 1$M_{\odot}$ ZAMS star of age $t = 5$ Gyr (see text for details). }
\label{fig:dynamics}
\end{figure*}

As the planet moves through the stellar atmosphere, the dissipation of shocks and turbulence creates a wake of hot gas behind it (see Fig.~\ref{fig:schematic} for a schematic illustration).  The temperature of this hot gas $T_{\rm h}$ is determined by equating its specific thermal energy $\sim c_{\rm s}^{2}$ with the power generated $\dot{E}/\dot{M} \sim (\mathcal{C}_{\rm d}/2)v_{\rm k}^{2}$ per rate of swept-up stellar mass $\dot{M} \simeq A_{\rm p}\rho_{\star}v_{\rm k}$, where $c_{\rm s}$ is the sound speed in the hot gas.  The evolution of $T_{\rm h}$ is shown in the bottom panel of Figure \ref{fig:dynamics}, indicates that $T_{\rm h}$ increases from $\sim 10^{6}$ K to $\sim 10^{7}$ K with increasing $\rho_{\star}$, i.e.~much hotter than the surface layers of the unperturbed star $T \sim 10^{4}-10^{5}$ K.

In order to calculate the emission from the stellar surface we must first estimate the size of the hot post-planetary layer.  The width of the layer perpendicular to the planetary motion $l_{\perp} \approx (R_{\rm p}H)^{1/2}$ is set by the width of the planet at the depth of the atmospheric scale height.  The length of the layer along the direction of motion $l_{\parallel} \sim t_{\rm cool}v_{\rm k}$ is instead set by the rate at which matter cools downstream, where $t_{\rm cool}$ is the cooling timescale.  

At early times, when the planet encounters the low-density outer layers of the star, the cooling timescale is set by the rate at which radiation diffuses vertically\footnote{The cooling and radiative diffusion timescales are equal because the energy density in the post-planetary wake is dominated by radiation.} $t_{\rm diff} \approx 3(H/c)\tau$, where $\tau \simeq \kappa_{\rm es}\rho_{\star}H$ is the vertical optical depth and $\kappa_{\rm es}$ is the opacity, the dominant form of which is Thomson scattering given the high temperatures $T_{\rm h} \gtrsim 10^{6}$ K of the shocked gas.  Although heating may substantially increase the scale height of the atmosphere further downstream from the planet (see below), we are justified in adopting an unperturbed value for the stellar scale height $H$ in calculating $t_{\rm diff}$ provided that cooling by diffusion is much faster than the rate of thermal expansion of the heated material (i.e. if $t_{\rm diff} \ll H/c_{\rm s}$), as is indeed satisfied when $\tau$ is sufficiently small.  

As the depth of the planet (and hence $\tau$ and $t_{\rm diff}$) increases, convection becomes more efficient than radiation at transporting energy generated by the planet to the stellar surface.  If convection dominates the cooling, then the length of the hot layer $l_{\parallel}$ can instead be determined by equating the dissipated power $\dot{E} \approx \dot{M}v_{\rm k}^{2}/2$ with the maximum luminosity that can be carried by convection $L_{\rm c,max} \approx (\rho_{\rm h}v_{\rm c,max}^{3}/2)\times (l_{\perp}l_{\parallel})$, where $\rho_{\rm h}$ is the density of the post-planetary gas and $v_{\rm c,max} = \alpha_{\rm c}c_{\rm s}$ (where $\alpha_{\rm c} \lesssim 1$) is the maximum convective velocity.  For the moment we assume that $\rho_{\rm h} \sim \rho_{\star}$, i.e.~that the gas in the hot layer does not have time to expand appreciably from its unperturbed value (and hence is drastically out of pressure equilibrium; we discuss the validity of this assumption below).  In this case the length of the hot layer is $l_{\parallel} \sim H/\alpha_{\rm c}^{3} \sim$ few $H$, corresponding to a convective cooling timescale $t_{\rm cool} \equiv t_{\rm c,vert}\sim$few $\times H/v_{\rm k}$.

Combining results, the emitting area of the hot post-planetary layer is approximately given by
\be
A_{\rm h} \simeq l_{\perp}\times l_{\parallel} \approx R_{\rm p}^{1/2}H^{1/2}v_{\rm k}t_{\rm cool},
\label{eq:Ah}
\ee
where $t_{\rm cool} \approx $ min[$t_{\rm diff} = 3H\tau_{\rm es}/c, t_{\rm c,vert} = 3H/v_{\rm k}$].  The transition between radiative and convective cooling ($t_{\rm diff} \gtrsim t_{\rm c,vert}$) occurs at stellar densities $\rho_{\star} \sim 10^{-5}$ g cm$^{-3}$ and is marked by the vertical dotted line in the bottom panel of Figure \ref{fig:dynamics}. 

Our estimate of $t_{\rm c,vert}$ above assumes that hot gas behind the planet does not expand appreciably from its unperturbed value, resulting in a hot wake of length $l_{\parallel} \sim$few $\times H$ when $t_{\rm diff} \gg t_{\rm c,vert}$.  In reality, the full extent of the heated region may be significantly larger than this estimate because the cooling timescale $t_{\rm c,vert} \propto \rho_{\rm h}^{-1}$ increases as the atmosphere expands, a process which itself occurs on a timescale $\sim H/c_{\rm s}$ similar to the cooling time.  Furthermore, in $\S\ref{sec:outflows}$ below we show that the hot post-planetary gas powers an outflow with a velocity $\sim c_{\rm s} \sim v_{\rm k}$, which itself acts to cool the gas on a similar timescale $\sim H/c_{\rm s}$.  

Because the timescales for convective cooling $t_{\rm c,vert}$, atmospheric expansion, and wind cooling are of the same order,\footnote{In principle cooling may also occur non-radiatively due to the redistribution of heat {\it latitudinally} by convective motions or subsonic excavation flows (e.g.~\citealt{Melosh89}), since in this case thermal energy is lost to PdV work via sideways expansion.  However, the condition for latitudinal transport to dominate over vertical transport is that $l_{\perp} \gtrsim H$, i.e. $R_{\rm p} \gtrsim H$, which is in practice only satisfied when the atmospheric density is sufficiently high $\rho_{\star} \gtrsim 10^{-2}$ g cm$^{-3}$ that the planet is already sinking dynamically (i.e.~$t_{\rm in,drag} \sim t_{\rm orb}$).} it seems reasonable to assume that a modest fraction $\epsilon_{\rm rad}$ of the generated thermal energy will transported to the stellar surface and radiated just a few scale heights behind the planet (see Fig.~\ref{fig:schematic}), resulting a luminosity $L_{\rm EUV/X} = \epsilon_{\rm rad}\dot{E}$.  Below, we adopt a fiducial value of $\epsilon_{\rm rad} = 0.1$, since obtaining a more accurate estimate of $\epsilon_{\rm rad}$ will require a detailed radiation-hydrodynamical simulation of the interaction between the planet and stellar atmosphere, a task beyond the scope of this paper.  In what follows below it should be kept in mind that the value of $\epsilon_{\rm rad}$ is highly uncertain (by at least an order of magnitude).  

The temperature of the emission from the hot post-planetary layer is approximately given by 
\begin{eqnarray} 
T_{\rm eff} &\simeq& \left(\frac{L}{A_h\sigma_{\rm sb}}\right)^{1/4} \nonumber \\ &\approx& 5\times 10^{5}{\,\rm K}\left(\frac{\epsilon_{\rm rad}}{0.1}\right)^{1/4}\left(\frac{\rho_{\star}}{10^{-3}\,\rm g\,cm^{-3}}\right)^{1/4}\left(\frac{H}{0.01R_{\odot}}\right)^{-1/8}
\label{eq:Teff}
\end{eqnarray}
where $\sigma_{\rm sb}$ is the Stefan-Boltzmann constant and we have used equation (\ref{eq:Ah}) for $A_{\rm h}$.  The evolution of $T_{\rm eff}$ is shown in the bottom panel of Figure \ref{fig:dynamics} assuming $\epsilon_{\rm rad} = 0.1$.  If the spectrum is approximately that of a blackbody, then the peak photon energy is $E_{\rm peak} \approx 3kT_{\rm eff} = 260(T_{\rm eff}/10^{6}$ K) eV.  Note that $T_{\rm eff}$ increases from an initial value of $\sim$few $10^{5}$ K (extreme ultraviolet; EUV), to a value $T_{\rm eff} \sim 10^{6}$ K (soft X-rays) at later times closer to merger.  In $\S\ref{sec:radiation}$ below we describe under what conditions this radiation escapes the vicinity of the star and hence is actually observable.

\subsection{Outflows}
\label{sec:outflows}

\begin{figure}
\resizebox{\hsize}{!}{\includegraphics[angle=0]{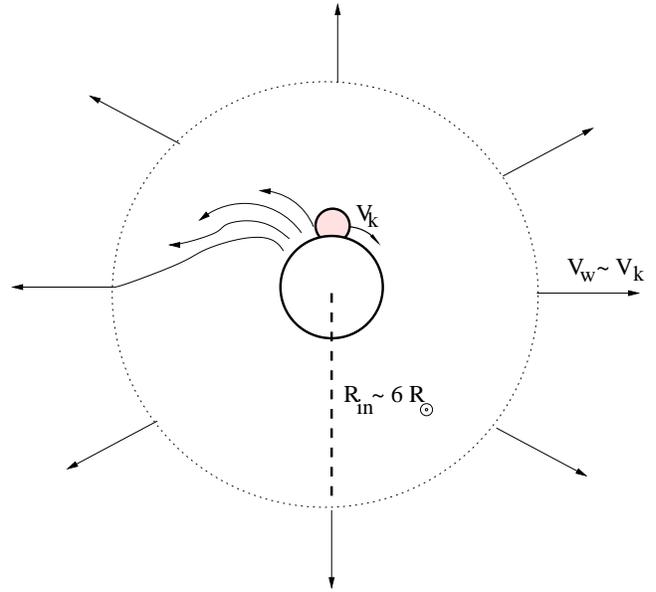}}
\caption[] {Schematic diagram of the outflow driven from the stellar surface by the energy generated as the planet grazes the stellar atmosphere.  The mass loss rate of the wind $\dot{M}_{\rm w}$ is comparable to the rate at which mass is swept up by the planet (eq.~[\ref{eq:Mdotw}]), whereas the asymptotic velocity of the wind is assumed to be similar to the stellar escape speed $\sim v_{\rm k}$.  On small radial scales the wind is non-stationary since it is launched from a (time varying) region well-localized behind the passing planet.  However, on larger scales $r \gtrsim R_{\rm in} \sim v_{\rm w}t_{\rm orb} \sim 6R_{\odot}$ (where $t_{\rm orb}$ is the planet orbital period) the outflow is reasonably well approximated as a spherically symmetric, steady-state wind.}
\label{fig:schematicwind}
\end{figure}

An outflow is likely from the hot post-planetary layer because the total generated power $\sim \dot{E}$ (eq.~[\ref{eq:edot}]) greatly exceeds the (local) Eddington luminosity $L_{\rm edd} \approx GM_{\star}A_{\rm h}c/R_{\star}^{2}\kappa_{\rm es}$, with their ratio given by
\begin{eqnarray}
\frac{\dot{E}}{L_{\rm edd}} &=& \frac{M_{\rm p}\kappa_{\rm es}}{2A_{\rm h}}\frac{v_{r}}{c} \nonumber \\ &\approx& 0.5\kappa_{\rm es}\rho_{\star}R_{\star}(v_{\rm k}/c) \approx 2\times 10^{4}\left(\frac{\rho_{\star}}{10^{-3}\rm g\,cm^{-3}}\right),
\label{eq:eddratio}
\end{eqnarray}
where we have used equations (\ref{eq:vr}) and (\ref{eq:Ah}) for $v_{r}$ and $A_{\rm h}$, respectively.  Equation (\ref{eq:eddratio}) shows that $\dot{E}/L_{\rm edd}\gtrsim 10^{2}-10^{5}$ for the relevant range of stellar densities $\rho_{\star} \sim 10^{-5}-10^{-2}$ g cm$^{-3}$ encountered during inspiral.     

If the fraction $1 - \epsilon_{\rm rad} \sim \mathcal{O}(1)$ of the thermal power that is not radiated immediately behind the planet is used to power a wind further downstream, then the resulting energy per unit mass $\gtrsim c_{\rm s}^{2}$ is similar to that required to escape the potential well of the star $\sim v_{\rm k}^{2}$.  A fraction of the thermal energy will go into expanding a hot envelope (e.g.~\citealt{Tylenda05}; \citealt{Soker&Tylenda06}), while a comparable fraction will go into the kinetic energy of an outflow.  Assuming that the asymptotic velocity of the wind $v_{\rm w}$ is comparable to the escape speed of the star $\sim v_{\rm k}$, then the mass loss rate $\dot{M}_{\rm w}$ of the outflow will be comparable to the rate at which the stellar atmosphere is swept up by the planet, i.e.~
\begin{eqnarray}
\dot{M}_{\rm w} \sim A_{\rm p}\rho_{\star}v_{\rm k} \approx 7\times 10^{22}{\,\rm g\,s^{-1}}\left(\frac{\rho_{\star}}{10^{-3}\,\rm g\,cm^{-3}}\right)\left(\frac{H}{10^{-2}R_{\odot}}\right)^{3/2}.
\label{eq:Mdotw}
\end{eqnarray}
Note that $\dot{M}_{\rm w}$ is many orders of magnitude greater than the normal mass loss rate in the solar wind $\sim 10^{12}$ g s$^{-1}$.  

Since the bulk of the wind is launched from a narrow layer behind the passing planet, the outflow will be non-stationary and highly asymmetric near the stellar surface.  However, on larger scales, greater than the distance $R_{\rm in} \approx v_{\rm w}t_{\rm orb} \approx 6R_{\odot}$ that wind material traverses in an orbital time, the flow will develop a structure similar to that of a spherically symmetric, steady-state wind.  Figure~\ref{fig:schematicwind} shows a schematic illustration of the wind geometry, showing an asymmetric inner cavity ($r \ll R_{\rm in}$) which connects onto a steady state wind at larger radii $r \gg R_{\rm in}$.  At sufficiently large radii, the density profile of the wind is approximately given by
\begin{eqnarray}
\rho_{\rm w} &\simeq& \frac{\dot{M}_{\rm w}}{4\pi r^{2}v_{\rm w}} \nonumber \\
&\approx &7\times 10^{-7}\rho_{\star}\left(\frac{H}{10^{-2}R_{\odot}}\right)^{3/2}\left(\frac{r}{6R_{\odot}}\right)^{-2}\,\,\,\,(r \gtrsim R_{\rm in}),
\label{eq:rhow}
\end{eqnarray}
where we have used equation (\ref{eq:Mdotw}) for $\dot{M}_{\rm w}$.

If the material in the wind expands adiabatically, then its temperature $T_{\rm w}(r)$ will be a factor ($\rho_{\rm w}/\rho_{\star}$)$^{\beta}$ smaller than the surface temperature of the hot layer $T_{\rm h}$, where $\beta \sim 1/3-2/3$ for a gas with adiabatic index $\Gamma \sim 4/3-5/3$ (depending on whether the outflow is dominated by gas or radiation pressure).  Near the inner edge $r = R_{\rm in} \sim 6R_{\odot}$ of the steady-state wind, typical temperatures are in the range $T_{\rm w} \approx 10^{3.5}-10^{5.5}$ K for $\rho_{\star} \sim 10^{-5}-10^{-2}$ g cm$^{-3}$.

\subsection{Electromagnetic Radiation}
\label{sec:radiation}

\begin{figure*}
\resizebox{14cm}{!}{\includegraphics[angle=0]{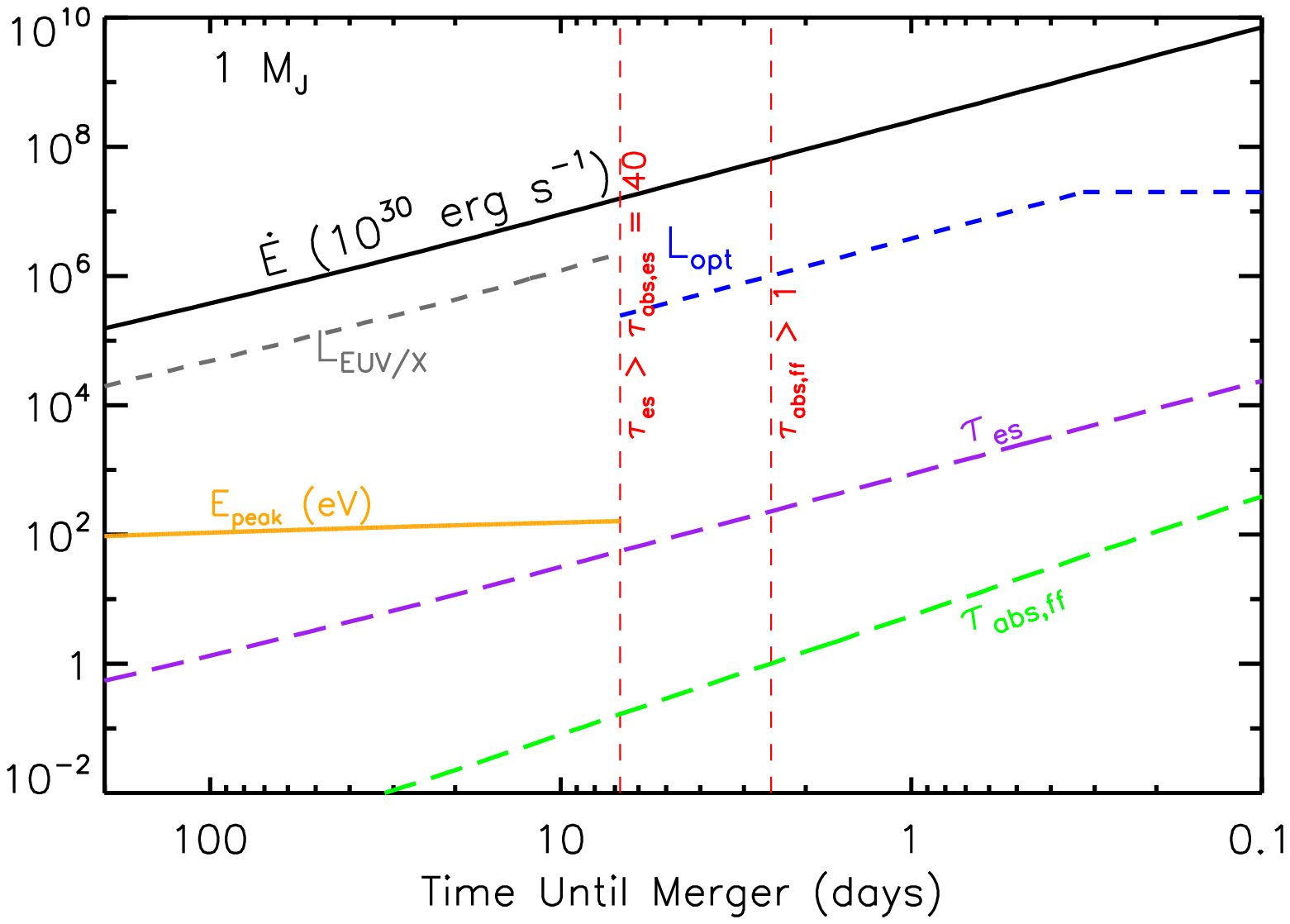}}\\
\resizebox{14cm}{!}{\includegraphics[angle=0]{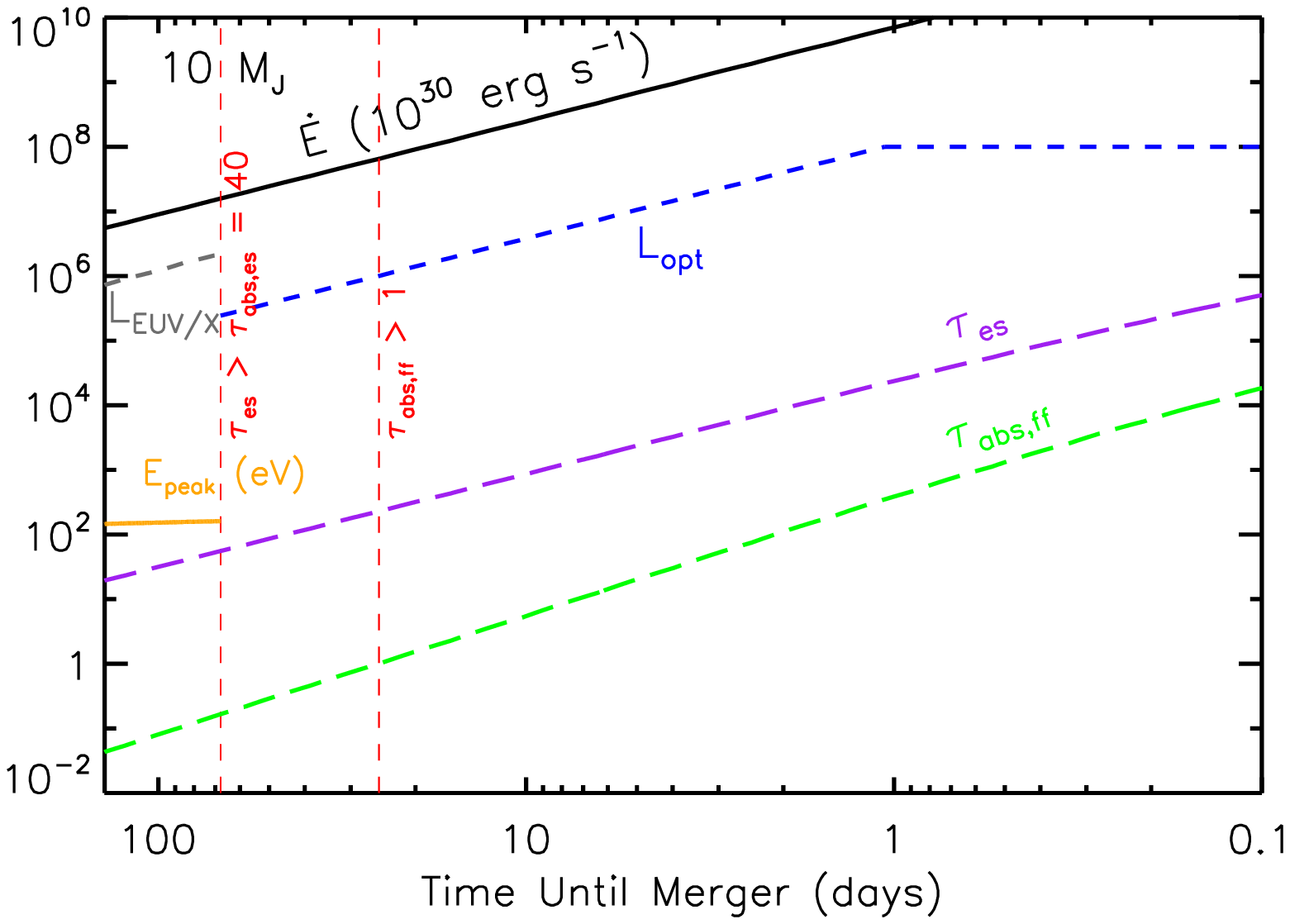}}
\caption[] {Properties of the radiation from a direct-impact planet-star merger, as a function of time until the dynamical plunge (``merger'') of the planet below the stellar surface $t_{\rm in,drag}$ (eq.~[\ref{eq:tin}]).  Top and bottom panels show cases corresponding to planet masses $M_{\rm p} = 1M_{\rm J}$ and $10M_{\rm J}$, respectively.  Quantities shown include the total rate at which energy is dissipated by the sinking planet $\dot{E}$ in units of 10$^{30}$ erg s$^{-1}$ ({\it solid black line}; eq.~[\ref{eq:edot}]); EUV/X-ray luminosity from the stellar surface just behind the hot planet in units of 10$^{30}$ erg s$^{-1}$ $L_{\rm EUV/X} \equiv \epsilon_{\rm rad}\dot{E}$ ({\it dashed grey line}; see $\S\ref{sec:Xray}$) assuming $\epsilon_{\rm rad} = 0.1$; peak photon energy from the hot layer behind the planet on the stellar surface $E_{\rm peak}$ ({\it orange line}; in eV); optical depth through the wind surrounding the star to electron scattering $\tau_{\rm es}$ ({\it dashed purple line}) and to free-free absorption $\tau_{\rm abs,ff}$ (at energies $E \approx E_{\rm peak}$ and accounting for the additional photon path-length due to electron scattering; {\it dashed green line}).  Vertical red dashed lines mark the times after which the photon luminosity from the stellar surface $L_{\rm EUV/X}$ is suppressed due to inelastic electron scattering ($\tau_{\rm es} \gtrsim 40$; $\tau_{\rm abs,es} > 1$) and free-free absorption ($\tau_{\rm abs,ff} > 1$).  A dashed blue line shows an estimate of the optical luminosity from the photosphere of the optically-thick stellar wind in units of 10$^{30}$ erg s$^{-1}$ ($\S\ref{sec:optical}$), which is estimated as $L_{\rm opt} =$ min[$L_{\rm rec},L_{\rm peak}$], where $L_{\rm rec}$ is the luminosity due to hydrogen recombination in the outflow (eq.~[\ref{eq:Lrec}]) and $L_{\rm peak}$ (eq.~[\ref{eq:Lpeak}]) is the peak luminosity achieved following the final inspiral stage, after which the outflowing material better resembles a dynamically ejected shell than a steady-state wind.}
\label{fig:radiation}
\end{figure*}

\subsubsection{EUV/X-ray Emission from the Early Inspiral}
\label{sec:Xray}

Figure \ref{fig:radiation} shows the power dissipated during inspiral for planet masses $M_{\rm p} = $1 and 10 $M_{\rm J}$ cases (top and bottom panels, respectively), similar to that shown in Figure \ref{fig:dynamics} but now plotted as a function of time until merger $t_{\rm in}$ (eq.~[\ref{eq:tin}]), starting at the time $t \sim 100$ days after which the infall rate due to gaseous drag exceeds that due to tidal dissipation.  Shown with a dark gray line is the (orbit-averaged) luminosity $L_{\rm EUV/X} = \epsilon_{\rm rad}\dot{E}$ from the hot post-planetary layer, assuming $\epsilon_{\rm rad} = 0.1$ (see above).  Shown in the bottom panel with an orange line is the peak photon energy $E_{\rm peak} = 3kT_{\rm eff}$ (eq.~[\ref{eq:Teff}]) in eV. 

Radiation from the stellar surface is in general not isotropic due to the geometry of the planet with respect to the hot radiating layer (see Fig.~\ref{fig:schematic}).  Rather, for a typical observer, emission is beamed into a fraction $f_{b} \equiv \Delta \Omega_{\rm rad}/4\pi \sim 0.25(H/R_{\rm p})^{1/2} \lesssim 0.1$ of the sky, as determined by the solid angle $\Delta \Omega_{\rm rad} \sim \pi(H/R_{\rm p})^{1/2}$ between the stellar surface and the planet.  Provided that radiation is free to propagate from the stellar surface to infinity unattenuated, one predicts that the observed emission will vary periodically on a timescale $t_{\rm orb}\sim$ hours, with an emission duty cycle $f_{b} \sim 10$ per cent and a peak luminosity $L_{\rm EUV/X}/f_{b} \approx 10L_{\rm EUV/X}$.  The shape of the light curve will depend both on the viewing angle of the observer and on the precise geometry of the radiating layer.  

If radiation indeed escapes from the stellar surface unattenuated, one would predict that $L_{\rm EUV/X}$ should increase from $\sim 10^{34}-10^{36}$ erg s$^{-1}$ at $t \sim$100 days prior to merger, up to $L_{\rm EUV/X} \sim 10^{41}-10^{42}$ erg s$^{-1}$ in the final hours before merger.  In reality, however, such a high luminosity during the final stages of the inspiral is not actually achieved because outflows from the stellar surface ($\S\ref{sec:outflows}$) block the escaping radiation once the mass loss rate becomes sufficiently high.  In particular, even if radiation escapes the vicinity of the stellar surface immediately behind the passing planet unhindered, it must necessarily pass through the mass lost from previous orbits (Figure \ref{fig:schematicwind}).  Since this ``orbit-averaged'' outflow is first encountered on a radial scale $\gtrsim R_{\rm in} \sim 6R_{\odot}$, the characteristic optical depth through the wind is $\tau \sim \rho_{\rm w}R_{\rm in}\kappa_{\rm w}$, where $\rho_{\rm w}$ (eq.~[\ref{eq:rhow}]) and $\kappa_{\rm w}$ are the density and opacity of the wind, respectively, at radii $r \sim R_{\rm in}$.

The opacity $\kappa_{\rm w}$ depends on the ionization state of the wind a radii $r \gtrsim R_{\rm in}$.  As described above, the mean energy of radiation from the stellar surface $E_{\rm peak} \gtrsim 100$ eV is sufficiently high $\gtrsim 13.6$ eV to ionize hydrogen.  Whether the wind is fully ionized thus depends on the ratio of the emission rate of ionizing photons $\dot{N}_{\rm ion} \sim L_{\rm EUV/X}/$Ry to the total rate of radiative recombination $\dot{N}_{\rm fb} \approx n_{\rm e}\sigma_{\rm fb}v_{\rm th}\times n_{\rm e}r^{3}$, viz.~
\begin{eqnarray}
\frac{\dot{N}_{\rm ion}}{\dot{N}_{\rm fb}} \sim 10^{3}\left(\frac{\epsilon_{\rm rad}}{0.1}\right)\left(\frac{\rho_{\star}}{10^{-3}\,\rm g\,cm^{-3}}\right)^{-1}\left(\frac{T_{\rm w}}{10^{4}\,\rm K}\right)^{1/2}\left(\frac{H}{10^{-2}R_{\odot}}\right)^{-3/2}\left(\frac{r}{6R_{\odot}}\right),
\label{eq:ndotratio}
\end{eqnarray}
where Ry = 13.6 eV is the Rydberg; $n_{\rm e} \approx \rho_{\rm w}/m_{\rm p}$ and $v_{\rm th} = (kT_{\rm w}/m_{\rm e})^{1/2}$ are the number density and thermal velocity of electrons in the wind, respectively; and $\sigma_{\rm fb} \approx 2\times 10^{-21}(T_{\rm w}/10^{4}{\,\rm K})^{-1}$ cm$^{2}$ is the radiative recombination cross section (Milne relation), where $T_{\rm w}$ is normalized to a typical value (see discussion after eq.~[\ref{eq:rhow}]).  Equation (\ref{eq:ndotratio}) shows that $\dot{N}_{\rm ion}/\dot{N}_{\rm fb}$ is $\gg 1$ for all relevant values of the stellar density and wind temperatures, indicating that the wind will be fully ionized at all radii $r \gtrsim R_{\rm in}$ as long as outflow is transparent to ionizing radiation, the conditions for which we now describe.

The two dominant sources of opacity in fully ionized matter are electron scattering and free-free absorption.  The electron scattering opacity $\kappa_{\rm es} \approx 0.4$ cm$^{2}$g$^{-1}$ is independent of frequency, whereas the free-free opacity $\kappa_{\rm ff} \propto \rho_{\rm w}T_{\rm w}^{1/2}\nu^{-3}$ on radial scale $\approx R_{\rm in}$ is given by
\begin{eqnarray}
\kappa_{\rm ff}|_{r \approx R_{\rm in}} &\approx&
10^{-5}{\,\rm cm^{2}\,g^{-1}}\left(\frac{\nu}{\nu_{\rm peak}}\right)^{-3}\left(\frac{\epsilon_{\rm rad}}{0.1}\right)^{-3/4}\times \nonumber \\
&&\left(\frac{\rho_{\star}}{10^{-3}\,\rm g\,cm^{-3}}\right)^{1/4}\left(\frac{H}{10^{-2}R_{\odot}}\right)^{11/8}\left(\frac{T_{\rm w}}{10^{4}\,\rm K}\right)^{-1/2},
\label{eq:kappaff}
\end{eqnarray}  
where we have normalized the incident frequency $\nu$ to the peak frequency $\nu_{\rm peak} = E_{\rm peak}/h \approx 3kT_{\rm eff}/h$ of the emission from the stellar surface using equation (\ref{eq:Teff}).  The low characteristic value of $\kappa_{\rm ff}$ in equation (\ref{eq:kappaff}) shows that, at the peak frequency, electron scattering dominates the opacity by several orders of magnitude (although free-free opacity becomes dominant at lower frequencies in the Rayleigh-Jeans tail). 

Figure \ref{fig:radiation} also shows the optical depth through the wind to electron scattering ({\it dashed purple line}).  When $\tau_{\rm es} \lesssim 1$, corresponding to $t_{\rm in} \gtrsim 100$ days, radiation escapes freely from the stellar surface, resulting in a periodic light curve with a peak luminosity $\gtrsim 10L_{\rm EUV/X}$ as described above.  On the other hand, at later times when $\tau_{\rm es} \gg 1$ scattering acts to effectively ``smear out'' this light curve variability.  Nevertheless, at lest initially the observed luminosity will continue to equal the orbit-averaged value $L_{\rm EUV/X}$ even when $\tau_{\rm es} \gtrsim 1$, because electron scattering is (to first order) energy conserving.

As the time until merger decreases, the wind becomes increasingly opaque, such that radiation from the stellar surface is completely absorbed.  The two dominant processes which thermalize photons in the wind are (1) inelasticity due to repeated electron scattering; and (2) free-free absorption.  Since the number of scattering events experienced by a photon is $\sim \tau_{\rm es}^{2}$ (for $\tau_{\rm es} > 1$), while a fraction $f_{sc} \sim h\nu_{\rm peak}/m_{\rm e}c^{2} \sim 5\times 10^{-4}$ of the original energy is lost per scattering, inelastic electron scattering results in appreciable energy loss once $\tau_{\rm es} \gtrsim f_{sc}^{-1/2} \equiv \tau_{abs,es} \approx 40$.  Figures \ref{fig:radiation} shows that for $M_{\rm p} = 1(10)M_{\rm J}$ this condition (marked by a vertical dashed line) is satisfied at times $t_{\rm in} \lesssim 7(70)$ days, when the EUV/X-ray luminosity is $L_{\rm EUV/X} \sim (\epsilon_{\rm rad}/0.1)10^{36}$ erg s$^{-1}$.  

Surface radiation is also attenuated by free-free absorption, which becomes more effective than suggested by equation (\ref{eq:kappaff}) due to the longer path length created by multiple electron scatterings.  One can account for this by defining an effective mean free path for thermal absorption as $\tau_{\rm abs,ff} = \tau_{\rm es}(\kappa_{\rm ff}/\kappa_{\rm es})^{1/2}$ (\citealt{Rybicki&Lightman79}), where $\kappa_{\rm ff}$ is the free-free opacity at the peak frequency (eq.~[\ref{eq:kappaff}]).  A second vertical dashed line Figure \ref{fig:radiation} marks the infall time below which free-free absorption is important ($\tau_{\rm abs,ff} > 1$), indicating that this process is somewhat less effective than inelastic electron scattering at thermalizing radiation near the peak of the EUV/X-ray spectrum. 

High-energy photons absorbed by the plasma are re-emitted as lower energy photons, for which the the free-free opacity is higher (e.g.~$\kappa_{\rm ff} \propto \nu^{-3}$; eq.~[\ref{eq:kappaff}]).  Once the timescale for photons to diffuse out of the wind becomes longer than the expansion timescale $\sim r/v_{\rm w}$, most of this energy is lost to PdV work.  Directly observable emission from the stellar surface is thus be substantially less than $L_{\rm EUV/X} = \epsilon_{\rm rad}\dot{E}$ after times for which $\tau_{\rm abs} \equiv$ min($\tau_{\rm abs,es}$,$\tau_{\rm abs,ff}$) $> 1$, as Figure \ref{fig:radiation} shows is satisfied in the weeks$-$months prior to merger.  The EUV/X-ray luminosity should thus peak at a value $L_{\rm EUV/X}(\tau_{\rm abs} \sim 1) \sim (\epsilon_{\rm rad}/0.1)10^{36}$ erg s$^{-1}$, after which the emission is strongly suppressed as the region surrounding the star becomes increasingly enshrouded by the merger-driven outflow.

\subsubsection{Optical Emission from the Final Inspiral}
\label{sec:optical}

In the previous section we showed that hard emission from the stellar surface is suppressed by absorption long before the final merger.  However, lower frequency emission from the outflow itself becomes visible, and continues to strengthen, as the outflow power approaches its peak value $\dot{E} \sim 10^{40}-10^{41}$ erg s$^{-1}$.  Most of the initial thermal energy in the wind at the stellar surface is lost to PdV expansion before it can be radiated at the photosphere of the wind.  However, the recombination of hydrogen contributes an additional source of luminosity (e.g.~\citealt{Kasen&Ramirez-Ruiz10}) and acts to slow the rate at which the temperature decreases (effectively reducing the adiabatic index).  The energy released by hydrogen recombination, if efficiently radiated, produces a `luminosity' given by
\be 
L_{\rm rec} \simeq (\dot{M}_{\rm w}/m_{\rm p})\times{\rm Ry} \sim \frac{\rm Ry}{m_{\rm p}v_{\rm k}^{2}}\dot{E} \sim 0.03\dot{E}.
\label{eq:Lrec}
\ee
Recombination is possible in the outflow only once energetic photons from the stellar surface are no longer able to keep the wind ionized; Figure \ref{fig:radiation} shows that this condition ($\tau_{\rm abs} \gtrsim 1$) occurs on a timescale of weeks$-$months prior to merger, when the outflow power $\dot{E} \sim 10^{37}$ erg s$^{-1}$ and the recombination luminosity is $L_{\rm rec} \sim 3\times 10^{35}$ erg s$^{-1}$. 

In $\S\ref{sec:exterior}$ we calculate the detailed properties of photospheric emission from outflows with very similar properties to those considered here ($\S\ref{sec:outflows}$), but powered instead by super-Eddington accretion following the tidal-disruption of a planet (see Figs.~\ref{fig:windprofile}, \ref{fig:superEdd}).  These calculations show that the radiated luminosity is indeed a significant fraction of the recombination luminosity $\sim L_{\rm rec}$ (eq.~[\ref{eq:Lrec}]), and that the radiation peaks at optical wavelengths (photospheric temperature $\sim 4-6\times 10^{3}$ K).  

Our results in $\S\ref{sec:exterior}$ show that equation (\ref{eq:Lrec}) provides a reasonable estimate of the optical luminosity during the earliest stages of the final inspiral $t_{\rm in,drag} \sim$ weeks$-$months.  However, equation (\ref{eq:Lrec}) overestimates the luminosity when the infall approaches its dynamical stage and $\dot{E}$ becomes very high.  This is because the assumption that the outflow is in steady state is no longer valid once the planetary infall time $t_{\rm in,drag}$ (eq.~[\ref{eq:tin}]) becomes shorter than the expansion timescale of the wind $t_{\rm exp} = R_{\rm rec}/v_{k}$ at the recombination radius $R_{\rm rec}$.  Since this condition $t_{\rm exp} > t_{\rm in,drag}$ is satisfied on a timescale $\sim$ day$-$week prior to merger, equation (\ref{eq:Lrec}) is only a valid estimate of the optical luminosity until the optical luminosity $L_{\rm opt}$ reaches a value $\sim 10^{37}-10^{38}$ erg s$^{-1}$.  

A significant fraction of the energy released, and total mass ejected, during the merger occurs during the final few orbits before the dynamical plunge.  An accurate treatment of the emission during this final phase when $t_{\rm exp} > t_{\rm in,drag}$ requires modeling the ejected mass as that of a discrete `shell' of ionized material with a mass\footnote{This factor 1/10 results because the planet falls a distance comparable to its own radius $\sim R_{\rm p} \sim 0.1R_{\star}$ in the final plunge, resulting in sufficient gravitational energy released to eject $\sim 1/10$ of its own mass.} $M_{\rm ej} \approx M_{\rm p}/10$.  The luminosity under this type of evolution does not peak on the same (short) timescale $\sim t_{\rm orb}$ of the final plunge (i.e.~when $\dot{E}$ peaks), but rather on the longer timescale set by when the expanding ejecta becomes sufficiently cool for hydrogen to recombine and radiate its energy.  

We may crudely estimate the peak luminosity $L_{\rm peak}$ and peak timescale $t_{\rm peak}$ of this final dynamical phase by equating $t_{\rm peak} \sim E_{\rm rec}/L_{\rm peak}$ with the timescale $t_{\rm exp} \sim R_{\rm photo}/v_{\rm k}$ required for the ejecta to expand to a sufficiently large radius $R_{\rm photo}$ for the energy released by recombination to be radiated, i.e.~$t_{\rm peak}L_{\rm peak} \approx E_{\rm rec}$, where $L_{\rm peak} \approx 4\pi \sigma_{\rm sb}T_{\rm rec}^{4}R_{\rm photo}^{2}$ is the thermal luminosity, $E_{\rm rec} \sim 0.1(M_{\rm p}/m_{\rm p})$Ry is the total energy released by recombination, and $T_{\rm rec} \approx 6000$ K is the temperature of the photosphere (set by the recombination of hydrogen; see Fig.~\ref{fig:windprofile}).  This results in the following analytic estimates:
\be
t_{\rm peak} \approx 1.3{\rm days}\left(\frac{T_{\rm rec}}{6000{\rm K}}\right)^{-4/3}\left(\frac{M_{\rm p}}{M_{\rm J}}\right)^{1/3}
\label{eq:tpeak}
\ee
\be
L_{\rm peak} \approx 2\times 10^{37}{\rm erg s^{-1}}\left(\frac{T_{\rm rec}}{6000{\rm K}}\right)^{4/3}\left(\frac{M_{\rm p}}{M_{\rm J}}\right)^{2/3}
\label{eq:Lpeak}
\ee
Thus, for $M_{\rm p} \sim 1-10M_{\rm J}$ the final merger is accompanied by an optical transient of luminosity $\sim 10^{37}-10^{38}$ erg s$^{-1}$ and characteristic duration $\sim$days (extending to times {\it after} the merger itself is complete).  As we describe further in $\S\ref{sec:detect}$ this event may resemble a classical nova, although is distinguished by its higher ejecta mass, lower ejecta velocity, and shorter typical duration.  In Figure \ref{fig:radiation} we plot the optical luminosity with a dashed blue line, assuming that $L_{\rm opt} \approx L_{\rm rec}$ at early times (when the approximation of a steady-state wind is valid), but then capping $L_{\rm opt} \approx L_{\rm peak}$ at late times when the evolution of the ejecta is dynamical.

\begin{figure*}
\resizebox{\hsize}{!}{\includegraphics[angle=0]{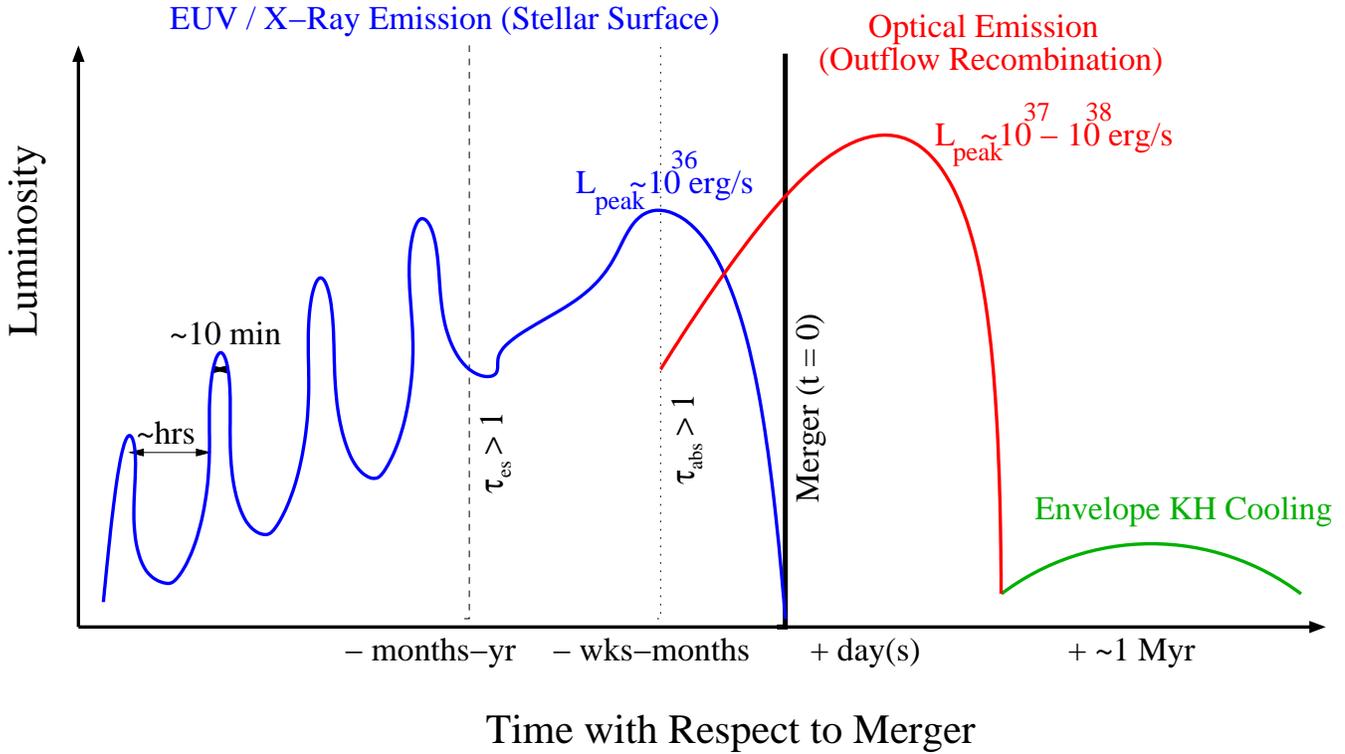}}
\caption[] {Schematic light curve of a direct-impact planet-star merger ($\S\ref{sec:interior}$).  At early times, EUV/X-ray emission ({\it blue}) originates from the hot layer behind the planet as it grazes the stellar atmosphere (see Fig.~\ref{fig:schematic}).  Initially the signal is periodic on the orbital timescale $t_{\rm orb} \sim$hours with a duty cycle $\sim 10$ per cent set by the narrow solid angle through which radiation escapes the cavity between the planet and the stellar surface.  The rate of atmospheric heating increases as the planet sinks deeper, creating an outflow from the stellar surface with a mass loss rate that increases with time (Fig.~\ref{fig:schematicwind}).  Electron scattering wipes out the X-ray periodicity months to a year prior to the merger, once the optical depth through the wind $\tau_{\rm es} \gtrsim 1$ ({\it dashed vertical line}).  Scattered ($\sim$isotropic) high energy radiation continues to escape for weeks$-$month prior to merger, before being entirely suppressed by other processes (inelastic electron scattering and free-free absorption) weeks$-$month prior to merger ($\tau_{\rm abs} \gtrsim 1$; {\it dashed vertical line}).  At late times, thermal optical emission from the photosphere of the wind ({\it red}) becomes increasingly bright, due in large part to the energy released by the recombination of hydrogen.  Although initially the optical luminosity traces the rate of mass loss from the wind, the final dynamical stages of the merger in effect results in the ejection of a shell of ionized material from the surface.  Thermal emission from this shell powers an optical transient lasting $\sim$day(s) after the merger, somewhat similar to a classical nova (see eqs.~[\ref{eq:tpeak}], [\ref{eq:Lpeak}]).  Finally, gravitational energy released by the planet deep inside the star is radiated on the much longer Kelvin-Helmholz cooling timescale of the stellar envelope $\sim 10^{6}$ yr ({\it green}).}
\label{fig:schematiclc}
\end{figure*}

Figure \ref{fig:schematiclc} shows a schematic diagram summarizing the various phases of light curve evolution for the case of direct-impact planetary mergers.

\section{Tidal-Disruption Event}
\label{sec:exterior}

If the planet undergoes Roche Lobe overflow above the stellar photosphere, then mass transfer may become unstable ($\S\ref{sec:outcomes}$; see Fig.~\ref{fig:demographics}).  In this case the planet is completely consumed in just a few orbital periods, resulting in a tidal-disruption event (see also \citealt{Bear+11}).  Since the Keplerian velocity greatly exceeds the sound speed of the planetary material, a shocked virialized torus is formed around the star with an angular momentum approximately equal to that of the original orbit.  A fraction of the disrupted planet will impact and mix with the stellar surface, whereas the majority will remain above the surface in rotationally-support configuration with characteristic radius $R_{d,0} \sim a_{\rm t} \gtrsim R_{\odot}$ (in some ways similar to the `excretion disks' observed around Be stars; e.g.~\citealt{Lee+91}).

The high temperature $T\gtrsim 10^{6}$ K in the disk implies that it will be fully ionized.  Turbulence generated by the magnetorotational instability (\citealt{Balbus&Hawley98}), or by shear instabilities in the layer between the disk and star, will then cause matter to accrete inwards at a characteristic rate
\begin{eqnarray}
\dot{M}_{0} &\sim& \frac{M_{\rm p}}{t_{\rm visc}} \nonumber \\
&\approx& 3\times 10^{25} {\rm\,g\,s^{-1}}\,\left(\frac{\alpha}{0.1}\right)\left(\frac{M_{\rm p}}{M_{\rm J}}\right)\left(\frac{M_{\star}}{M_{\odot}}\right)^{1/2}\left(\frac{R_{\rm d}}{R_{\odot}}\right)^{-3/2}\left(\frac{H_{\rm d}/r}{0.5}\right)^{2},\nonumber \\
\label{eq:mdot0}
\end{eqnarray} where 
\begin{eqnarray}
t_{\rm visc} &\simeq& t_{\rm orb}\alpha^{-1}\left(\frac{H_{\rm d}}{R_{t}}\right)^{-2} \nonumber \\ 
&\simeq& \frac{R_{t}^{2}}{\nu} \approx 7\times 10^{4}{\rm\,s}\left(\frac{\alpha}{0.1}\right)^{-1}\left(\frac{M_{\star}}{M_{\odot}}\right)^{-1/2}\left(\frac{R_{\rm d}}{R_{\odot}}\right)^{3/2}\left(\frac{H_{\rm d}/r}{0.5}\right)^{-2}
\label{eq:tvisc}
\end{eqnarray}
is the viscous accretion timescale (typically $\sim$days-weeks); $\nu = \alpha c_{\rm s}H_{\rm d}$ is the effective kinematic viscosity; $c_{\rm s} = H_{\rm d}\Omega$ is the midplane sound speed in vertical hydrostatic equilibrium; $H_{\rm d}$ is the vertical scale height of the disk; $\Omega = (GM_{\star}/r^{3})^{1/2}$ is the Keplerian orbital frequency; and $\alpha \sim 0.01-0.1$ is the parametrized dimensionless viscosity (\citealt{Shakura&Sunyaev73}).   

Equation (\ref{eq:mdot0}) shows that the initial accretion rate $\dot{M}_{0}$ is typically $\sim 10^{2}-10^{4}$ times higher than the Eddington accretion rate $\dot{M}_{\rm edd,\star} \simeq 7\times 10^{22}(M_{\star}/M_{\odot})(R_{\star}/R_{\odot})$ g s$^{-1}$ at the stellar surface, depending on the values of $M_{\rm p}$ and $\alpha$.  At such high accretion rates the disk cannot cool efficiently and is thus geometrically thick with $H_{\rm d} \sim R_{\rm d}/2$.  This is because the timescale for photons to diffuse out through the midplane 
\be
t_{\rm diff} \simeq (3H_{\rm d}/c)\tau_{\rm d} \approx 6\times 10^{7}{\,\rm s}\left(\frac{M_{\rm p}}{M_{\rm J}}\right)\left(\frac{R_{\rm d}}{R_{\odot}}\right)^{-1}\left(\frac{H_{\rm d}/r}{0.5}\right)
\ee 
is much longer than the accretion timescale $t_{\rm visc}$ (eq.~[\ref{eq:tvisc}]) over which gravitational energy is released, where $\tau_{\rm d} \simeq \rho_{\rm d}H_{\rm d}\kappa_{\rm es}$ is the vertical optical depth, $\rho_{\rm d} = M_{\rm p}/2\pi H_{\rm d}R_{\rm d}^{2}$ is the midplane density, and $\kappa_{\rm es}$ is the electron scattering opacity.  More generally, the condition $t_{\rm diff} \gtrsim t_{\rm visc}$ is satisfied at radii less than the critical ``trapping radius''
\be
R_{\rm tr} = \frac{\dot{M}\kappa_{\rm es}}{4\pi c} \simeq R_{\star}\left(\frac{\dot{M}}{\dot{M}_{\rm Edd,\star}}\right)
\label{eq:rtrap}
\ee

For a thick disk with $H_{\rm d} \sim R_{\rm d}$ that evolves due to the viscous redistribution of angular momentum, the accretion rate onto the star evolves with time approximately as (e.g.~\citealt{Pringle81}; \citealt{Metzger+08})
\begin{eqnarray}
\dot{M}(t) \simeq
\left\{
\begin{array}{lr}
 \dot{M}_{0}
, &
t \le t_{\rm visc} \\
\dot{M}_{0}(t/t_{\rm visc})^{-4/3},&
t > t_{\rm visc} \\
\end{array}
\right.. 
\label{eq:mdot}
\end{eqnarray}
If the evolution of the disk conserves total angular momentum\footnote{In reality, a fraction of the original angular momentum of the orbit will go into the outer layers of the star rather than into the disk.  Nevertheless, the self-similar evolution of the disk radius $R_{\rm d} \propto t^{2/3}$ should still obtain at late times, although the normalization of $R_{\rm d}$ may be somewhat smaller than our estimate in equation (\ref{eq:Rd}).} $J_{\rm d} \propto M_{\rm d}(GM_{\star}R_{\rm d})^{1/2}$, where $M_{\rm d}$ and $R_{\rm d}$ are the total mass and (mass-)averaged disk radius, then the outer edge of the disc viscously spreads outwards in time as $R_{\rm d} \propto M_{\rm d}^{-2}$, i.e.~
 \begin{eqnarray}
R_{\rm d}(t) \simeq
\left\{
\begin{array}{lr}
 R_{\star}
, &
t \le t_{\rm visc} \\
R_{\star}(t/t_{\rm visc})^{2/3},&
t > t_{\rm visc} \\
\end{array}
\right.. 
\label{eq:Rd}
\end{eqnarray}
Equations (\ref{eq:mdot}) and (\ref{eq:Rd}) do not strictly apply once the outer edge of the disk is able to cool efficiently into a thin disk (i.e.~once $R_{\rm d} \gtrsim R_{\rm tr}$).  These expression nevertheless provide a reasonable approximation to the viscous evolution predicted by more realistic models (e.g.~\citealt{Cannizzo+90}).\footnote{For a geometrically thin $\alpha-$disk dominated by gas pressure and Kramers opacity, we find that $\dot{M} \propto t^{-1.25}$ and $R_{\rm d} \propto t^{1.5}$ at times $t \gg t_{\rm visc}$.}

The timeline of the disk evolution can thus be summarized as follows.  At early times after the disk forms, photons are trapped throughout the entire disk, i.e.~ $R_{\rm tr} \gg R_{\rm d}$ ($\dot{M} \gg \dot{M}_{\rm edd,\star}$).  Since $R_{\rm tr}/R_{\rm d} \propto t^{-2}$, as time proceeds the outer edge of the disk evolves to become `untrapped' and radiatively efficient (i.e.~$R_{\rm d} \gtrsim R_{\rm tr}$) on a timescale $\sim (\dot{M}_{0}/\dot{M}_{\rm Edd,\star})^{1/2}t_{\rm visc} \sim$month when $R_{\rm tr} \approx R_{\rm d} \approx 10R_{\star}$.  After this time, the trapping radius continues to move inwards, reaching the stellar surface at a time 
\begin{eqnarray}
t_{\rm edd} \approx t_{\rm visc}\left(\frac{\dot{M}_{0}}{\dot{M}_{\rm Edd,\star}}\right)^{3/4} \approx 80{\rm \,days\,}\left(\frac{\alpha}{0.1}\right)^{1/4}\left(\frac{M_{\rm p}}{M_{\rm J}}\right)^{3/4}\left(\frac{M_{\star}}{M_{\odot}}\right)^{7/8}.
\label{eq:tedd}
\end{eqnarray}
At this point the outer radius of the disk is located at
\begin{eqnarray}
R_{\rm edd} &\equiv& R_{\rm edd}(t=t_{\rm edd}) \nonumber \\ &\approx& R_{d,0}\left(\frac{\dot{M}_{0}}{\dot{M}_{\rm Edd,\star}}\right)^{2/3}\approx 60R_{\star}\left(\frac{\alpha}{0.1}\right)^{2/3}\left(\frac{M_{\rm p}}{M_{\rm J}}\right)^{2/3}\left(\frac{M_{\star}}{M_{\odot}}\right)^{1/3}.
\label{eq:Redd}
\end{eqnarray}
Thus, depending on the value of $\alpha \sim 0.01-0.1$ and $M_{\rm p} \sim 1-10M_{\rm J}$ we expect $t_{\rm edd} \approx $month-few years and $R_{\rm edd} \approx 10-300R_{\odot}$.

In the following sections we estimate the properties of the thermal radiation from the accretion disk at times $t < t_{\rm edd}$ ($\S\ref{sec:superEdd}$) and $t > t_{\rm edd}$ ($\S\ref{sec:subEdd}$).

\subsection{Emission from Super-Eddington Outflows ($t < t_{\rm edd}$)}
\label{sec:superEdd}

\begin{figure}
\resizebox{\hsize}{!}{\includegraphics[angle=0]{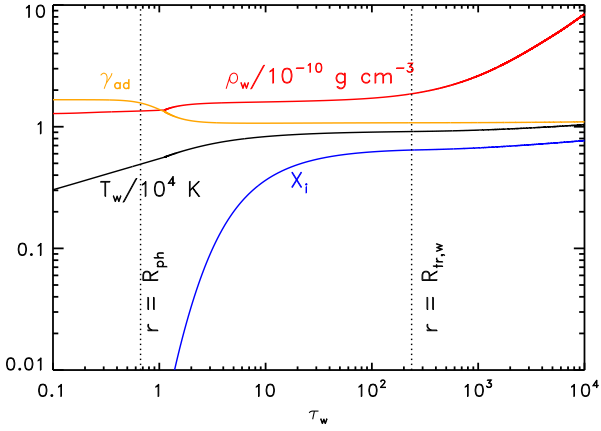}}
\caption[] {Example solution showing the structure of an outflow powered by super-Eddington accretion following the tidal-disruption of a Jupiter-mass planet, as a function of the optical depth from the surface $\tau_{\rm w}$.  The solution is shown at a time $t = t_{\rm visc} \approx 1.5$ days, when the outflow velocity is $v_{\rm w} \approx 400$ km s$^{-1}$ and the radiated luminosity is $L \approx 1.2 \times 10^{37}$ erg s$^{-1}$ (see Figure \ref{fig:superEdd}).  The disk is assumed to have viscosity $\alpha = 0.1$ and initial radius $R_{d,0} = 1.5R_{\odot}$.  Properties shown include the wind temperature $T_{\rm w}$ ({\it black line}); wind density $\rho_{\rm w}$ ({\it red line}); ionized fraction of hydrogen $X_{i}$ ({\it blue line}); and effective adiabatic index $\gamma_{\rm ad}$ ({\it orange line}).  Also shown with black vertical dashed lines are the wind trapping radius $R_{\rm tr,w} \equiv r(\tau_{\rm w} = c/3v_{\rm w})$ and the photospheric radius $R_{\rm ph} \equiv r(\tau_{\rm w} = 2/3)$.}
\label{fig:windprofile}
\end{figure}

\begin{figure}
\centering
\subfigure[$M_{\rm p} = 1M_{\rm J}$]{
\includegraphics[width = 0.48\textwidth]{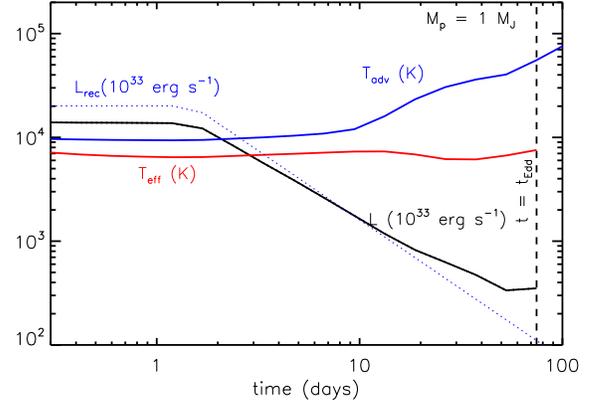}}
\subfigure[$M_{\rm p} = 10M_{\rm J}$]{
\includegraphics[width = 0.48\textwidth]{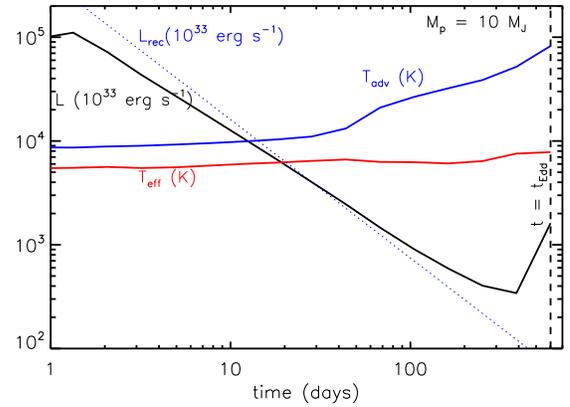}}
\caption[]{Properties of thermal emission from accretion disk outflows, as a function of time since planetary disruption, during the super-Eddington accretion phase ($t < t_{\rm Edd}$; eq.~[\ref{eq:tedd}]).  Top and bottom panels show cases corresponding to a disrupted planet of mass $M_{\rm p} = 1M_{\rm J}$ and $10 M_{\rm J}$, respectively.  The wind mass loss rate is assumed to be a fraction $f_{\rm w} = 0.1$ of the accretion rate $\dot{M}$ (eq.~[\ref{eq:mdot}]), while the thermal energy at the base of the wind is assumed to be a fraction $\epsilon_{\rm th} = 0.1$ of the outflow kinetic energy.  Quantities show include the radiated bolometric luminosity in units of $10^{33}$ erg s$^{-1}$ ({\it solid blue line}); effective temperature of radiation at the photosphere $T_{\rm eff} \equiv T_{\rm w}(\tau_{\rm w} = 2/3$) ({\it solid red line}); and temperature of the wind at the ``advection'' radius $T_{\rm adv} \equiv T_{\rm w}(\tau_{\rm w} = c/3v_{\rm w}$) ({\it solid red line}).  Shown for comparison is the maximum luminosity due to the recombination of hydrogen $L_{\rm rec}$ (eq.~[\ref{eq:Lrec}]).  Note that at early times the radiated luminosity is a significant fraction of $L_{\rm rec}$, while at late times $L > L_{\rm rec}$ once the photosphere recedes to near the outflow surface, such that thermal energy advected from the base of the wind contributes most of the luminosity.}
\label{fig:superEdd}
\end{figure}

At early times $t \lesssim t_{\rm edd}$ when the accretion rate is highly super-Eddington, powerful outflows are likely to occur driven by radiation pressure, either directly from the accretion disk (e.g.~\citealt{Ohsuga+05}) or from the stellar surface, which is heated by energy released at the shear interface between the disk and star (as in the case of direct merger; $\S\ref{sec:interior}$).  Numerical simulations (e.g.~\citealt{Ohsuga+05}) and observations of super-Eddington accreting systems (e.g.~SS433; \citealt{Begelman+06}) both suggest that a substantial fraction of the accreting mass is lost in outflows, such that the wind mass loss rate is $\dot{M}_{\rm w} \approx f_{\rm w}\dot{M}$ with $f_{\rm w} \sim \mathcal{O}(1)$.  If one assumes that the total gravitational energy released as planetary material settles from the point of disruption to the stellar surface goes into the kinetic energy of the outflow, we estimate that the maximum value of $f_{\rm w}$ is $\sim 0.3-0.5$ for planet-to-star density ratios in the range $\bar{\rho}_{\rm p}/\bar{\rho}_{\star} \sim 1-5$ corresponding to unstable mass transfer; hereafter we adopt a fiducial value of $f_{\rm w} = 0.1$.

Because the outflow is highly optically thick, emission from the accretion disk itself is not visible at these early times.  Nevertheless, bright thermal emission can originate from the photosphere of the wind, as has been studied before (e.g.~\citealt{Rossi&Begelman09}; \citealt{Strubbe&Quataert09}) in the context of other super-Eddington systems.  In this section we calculate the light curve and temperature of the emission from these super-Eddington outflows.

We assume that the wind is launched from a characteristic radius $R_{\rm w}(t) \simeq$ min[$R_{\rm tr},R_{\rm d}] \sim 1-10R_{\odot}$ set by the maximum radius where the disk is locally super-Eddington, where $R_{\rm tr}(t)$ and $R_{\rm d}(t)$ are calculated using equations (\ref{eq:rtrap}) and (\ref{eq:Rd}), respectively.   We furthermore assume that the asymptotic speed of the wind approximately equals the escape speed from the launching radius $v_{\rm w}(t) \approx (GM_{\star}/R_{\rm t})^{1/2} \approx 100-400$ km s$^{-1}$, as is a common property of thermal pressure-driven winds (e.g.~\citealt{Lamers&Cassinelli99}).  At most times the accretion rate (and hence the wind mass loss rate) evolves on a timescale that is relatively slow compared to the time since disruption (see eq.~[\ref{eq:mdot}]) or the expansion timescale at the photosphere.  In this case it is reasonable to approximate the density profile of the outflow as that of a steady-state wind:
\be
\rho_{\rm w}(r,t) \simeq \frac{\dot{M}_{\rm w}(t)}{4\pi r^{2}v_{\rm w}}\exp[-(r/r_{\rm edge})^{2}],
\ee
where the exponential cut-off accounts for the outer edge of the ejecta at $r \sim r_{\rm edge} = v_{\rm w}t$.  The optical depth through the wind to radius $r$ at any time $t$ is given by
\be
\tau_{\rm w}(r,t) \approx \int_{r}^{\infty}\rho_{\rm w}\kappa_{\rm w}dr',
\ee
where $\kappa_{\rm w}(\rho_{\rm w},T_{\rm w})$ is the Rosseland mean opacity and $T_{\rm w}(r)$ is the temperature profile of the wind.

At small radii in the wind (high $\tau_{w}$) we calculate $T_{\rm w}(r)$ by assuming that entropy is conserved from the base of the wind at $r = R_{\rm w}$, out to the wind `trapping' radius $R_{\rm tr,w}$ above which radiation freely escapes without further loss of energy.\footnote{Not to be confused with the trapping radius in the disk (eq.~[\ref{eq:rtrap}]).}  The latter is defined as the radius above which the expansion timescale in the wind $t_{\rm exp} \approx r/v_{\rm w}$ is greater than the timescale for outwards radiative diffusion $t_{\rm diff,w} \approx 3\tau_{\rm w}r/c$.  The thermal energy at the base of the wind is assumed to be a fraction $\epsilon_{\rm th} \lesssim 1$ of the wind kinetic energy $\rho_{\rm w}v_{\rm w}^{2}/2$.  The value of $\epsilon_{\rm th}$ is highly uncertain but should be of order unity if the outflow represents matter preferentially heated to virial temperatures, i.e.~above the entropy of the midplane.  We assume a value of $\epsilon_{\rm th} = 0.1$ in our calculations below, but our results do not depend sensitively on its precise value.

Outside the trapping radius the wind is effectively stationary on the expansion timescale and can thus be approximated as a radiative stellar atmosphere.  For $r > R_{\rm tr,w}$ the temperature gradient is thus given by (e.g.~\citealt{Hansen&Kawaler94})
\be
\frac{\partial T_{\rm w}}{\partial r} = -\frac{3}{16}\frac{L}{4\pi r^{2}}\frac{\kappa_{\rm w}\rho_{\rm w}}{\sigma_{\rm sb}T^{3}}\,\,\,(r > R_{\rm tr,w}).
\ee
Here $L$ is the photon luminosity of the wind.  Its initial value at the trapping radius is set by the advection of thermal energy across the trapping radius, i.e. $L(r = R_{\rm tr,w}) = (16\pi/3)\sigma_{\rm sb}T_{\rm w}^{4}r^{2}|_{r = R_{\rm tr,w}}\times (v_{\rm w}/c)$ (e.g.~\citealt{Rossi&Begelman09}).  At larger radii $L$ gains contributions from the additional energy input from hydrogen recombination, i.e. $dL_{\rm rec} = (\dot{M}_{\rm w}/m_{\rm p})|dX_{i}|\times$Ry where $|dX_{i}|$ is the change in the ionized fraction $X_{i}(r)$.

We calculate the ionization state $X_{i}(r)$ of the wind assuming local thermodynamic equilibrium (Saha equation), as in \citet{Kasen&Ramirez-Ruiz10}.  The opacity of the ejecta $\kappa_{\rm w}$ is determined using data from the Opacity Project\footnote{See\,\url{http://cdsweb.u-strasbg.fr/topbase/OpacityTables.html}}, assuming the metallicity of the gas to be five times solar (a reasonable approximation to the composition of gaseous giant planets).  The observed luminosity $L(t) = 4\pi r^{2}\sigma_{\rm sb}T_{\rm w}^{4}$ and effective temperature $T_{\rm eff}(t) = T_{\rm w}$ of the radiation are those near the photosphere of the outflow ($\tau_{\rm w} = 2/3$).  Note that since the trapping radius is a function of $\tau_{\rm w}(r)$, which depends on the temperature structure of the wind, the self-consistent structure of the wind must be solved by iteratively ``guessing'' the location of the trapping radius.  

Figure \ref{fig:superEdd} shows our calculation of the time evolution of the luminosity $L$; outflow temperature at the trapping radius $T_{\rm adv} \equiv T_{\rm w}(r = r_{\rm adv})$; and effective temperature $T_{\rm eff} \equiv (L/4\pi \sigma_{\rm sb}R_{\rm ph}^{2})^{1/4}$ of the emission, where $R_{\rm ph} \equiv r(\tau_{\rm w} = 2/3)$ is the radius of the photosphere.  In the top and bottom panels, respectively, we show cases corresponding to different assumed planetary masses $M_{\rm p} = 1M_{\rm J}$ and $M_{\rm p} = 10M_{\rm J}$.  In all calculations we assume a wind mass loss parameter $f_{\rm w} = 0.1$; disk viscosity $\alpha = 0.1$; and initial disk radius $R_{\rm d} = 1.5R_{\odot}$, resulting in an Eddington time of $t_{\rm edd} \sim 80(450)$ days for $M_{\rm p} = 1(10)M_{\rm J}$ (eq.~[\ref{eq:tedd}]).  For comparison we also show the maximum luminosity due to hydrogen recombination $L_{\rm rec}$ (eq.~[\ref{eq:Lrec}]) with a dashed blue line.

Figure \ref{fig:superEdd} shows that the wind luminosity peaks at a luminosity $L \sim 10^{37}(10^{38})$ erg s$^{-1}$ for $M_{\rm p} = 1(10)M_{\rm J}$, in both cases on a timescale $\sim t_{\rm visc} \sim 1$ day, before declining with time approximately as $L \propto t^{-1}$ (somewhat more shallow than $\dot{M} \propto t^{-4/3}$).  During this time the effective temperature is relatively constant at $T_{\rm eff} \approx 5000-7000$ K, due largely to the sensitive temperature dependence of the opacity in this temperature range.  At early times ($t \lesssim 10$ days) most of the luminosity results from the energy released by hydrogen recombination, i.e. $L \lesssim L_{\rm rec}$ (eq.~[\ref{eq:Lrec}]), as opposed to the initial thermal energy advected from the base of the wind.  This implies that the evolution near peak luminosity is relatively insensitive to the precise value assumed for $\epsilon_{\rm th}$.  On the other hand, at later times ($t \gtrsim 10$ days) the radiated luminosity exceeds the maximum power due to recombination alone ($L > L_{\rm rec}$).  This is because the thermal energy advected from the base of the wind becomes more important once the mass loss rate (and hence opacity) of the wind decreases sufficiently.  This transition is evident also in the late-time rise of the temperature at the advection radius $T_{\rm adv}$.

Figure \ref{fig:windprofile} shown an example of the wind structure with optical depth for the $M_{\rm p} = 1M_{\rm J}$ case from Figure \ref{fig:superEdd}, corresponding to a time $t = t_{\rm visc} \approx 1.5$ days after disruption when the radiated luminosity is $L \approx 1.2\times 10^{37}$ erg s$^{-1}$.  Note that the ionized fraction $X_{i} \approx 0.7$ is still relatively high at the trapping radius, but that recombination is essentially complete by the time matter reaches the photosphere ($X_{i} \approx 0$).  Since energy released exterior to the trapping radius escapes without adiabatic losses, this implies that a large fraction of the total recombination luminosity $L_{\rm rec}$ (eq.~[\ref{eq:Lrec}]) is radiated.  

We conclude that the early super-Eddington accretion phase of tidal-disruption planet-star mergers is accompanied by a bright optical transient with a peak duration $t_{\rm visc} \sim$ day-week, which is followed by a general dimming and blueward spectral shift lasting for a time $t_{\rm Edd} \sim $ months$-$year.  As we now describe, this dimming is only temporary, since emission re-brightens once radiation from the disk itself becomes visible following the super-Eddington phase.     

\begin{figure}
\resizebox{\hsize}{!}{\includegraphics[angle=0]{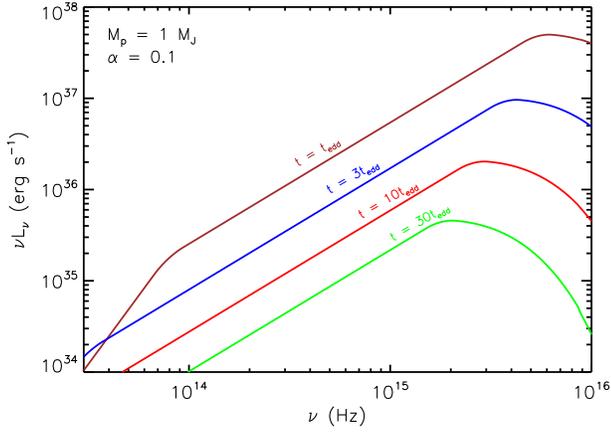}}
\caption[] {Spectra of thermal emission (multi-color blackbody) from the sub-Eddington accretion phase ($t \gtrsim t_{\rm edd}$; $\S\ref{sec:subEdd}$) following the tidal-disruption of a planet, shown at several times $t/t_{\rm edd} =$ 1 ({\it brown}),3 ({\it blue}),10 ({\it red}), and 30 ({\it green}), where $t_{\rm edd} = $80 days is the Eddington timescale (eq.~[\ref{eq:tedd}]).  The calculation shown assumes the mass of the disrupted planet $M_{\rm p} = 1M_{\rm J}$ and disk viscosity $\alpha = 0.1$.}
\label{fig:disknuLnu}
\end{figure}

\begin{figure}
\centering
\subfigure[$M_{\rm p} = 1M_{\rm J}$]{
\includegraphics[width = 0.48\textwidth]{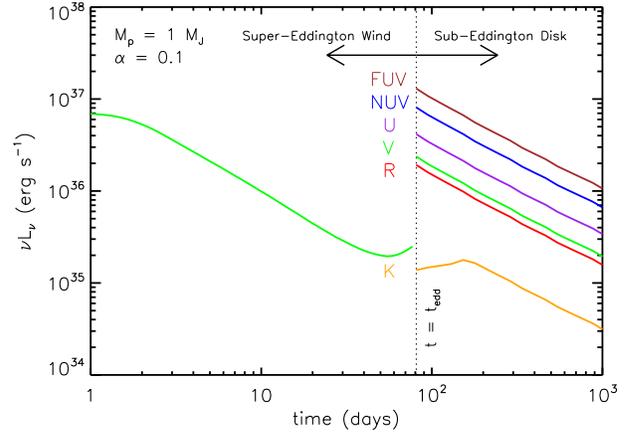}}
\subfigure[$M_{\rm p} = 10M_{\rm J}$]{
\includegraphics[width = 0.48\textwidth]{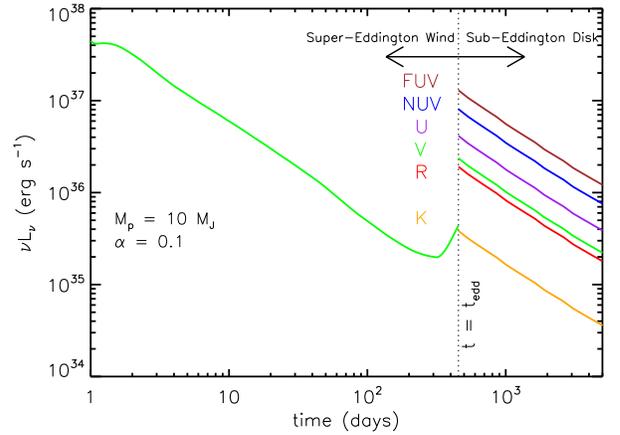}}
\caption[]{Light curves of thermal emission from the sub-Eddington accretion phase at various observed frequencies, including FUV ({\it brown}), NUV({\it blue}), U({\it green}), V({\it orange}), R({\it purple}), and K ({\it red}) bands.  Top and bottom panels show cases corresponding to a disrupted planet of mass $M_{\rm p} = 1M_{\rm J}$ and 10$M_{\rm J}$, respectively.  Shown for comparison at times $t < t_{\rm edd}$ is the visual emission from disk outflows during the preceding super-Eddington accretion phase ($\S\ref{sec:superEdd}$; see Fig.~\ref{fig:superEdd}).  The visual light curve discontinuity before and after the Eddington transition indicates that an intermediate transitional phase (not shown) will smoothly connect these phases.}
\label{fig:disklc}
\end{figure}

\subsection{Accretion Disk Emission (Sub-Eddington Phase; $t > t_{\rm edd}$)}
\label{sec:subEdd}

Powerful outflows from the disk cease once the accretion rate decreases below the Eddington rate on a timescale $t \gtrsim t_{\rm edd} \sim$months$-$year.  After this point emission from the accretion disk itself becomes visible,\footnote{In principle ejecta from the super-Eddington phase could still be opaque, even long after the outflow subsides.  However, a calculation similar to that presented in $\S\ref{sec:radiation}$ shows that photons from the inner disk will fully ionize the outflow, and that only a short period of free expansion is required for the ejecta to become optically thin to Thomson scattering.} with a bolometric luminosity $L \sim 10^{38}$ ergs s$^{-1}(\dot{M}/\dot{M}_{\rm edd,\star})$ that is comparable or greater than the peak luminosity during the super-Eddington phase (and much greater than the luminosity at times $t \lesssim t_{\rm edd}$; see Fig.~\ref{fig:superEdd}).  The sub-Eddington transition thus signals the onset of much brighter and harder emission.

Since the disk is geometrically thin when $\dot{M} \lesssim \dot{M}_{\rm edd}$, its spectrum can be reasonably well described as that of a multi-colored blackbody, which extends from the stellar surface $r \approx R_{\star}$ to the outer edge of the disk $r = R_{\rm d}(t)$ (eq.~[\ref{eq:Rd}]).  The effective temperature at any radius in the disk is given by equating the radiated flux $\sigma_{\rm sb}T_{\rm eff}^{4}$ with the gravitational power released $\simeq (3/8\pi)GM_{\star}\dot{M}_{\rm d}r^{-3}$.  The peak frequency of emission resulting from radius $r$ is thus given by
\be
h\nu_{\rm peak} \approx 3kT_{\rm eff} \approx 17{\rm eV}\left(\frac{\dot{M}_{\rm d}}{\dot{M}_{\rm edd}}\right)^{1/4}\left(\frac{r}{R_{\odot}}\right)^{-3/4},
\label{eq:hnupeak}
\ee
such that we expect the disk emission to range from ultraviolet (UV) to optical-infrared wavelengths. 

Figure \ref{fig:disknuLnu} shows our calculation of the spectrum of thermal disk emission for the case of a disrupted planet with mass $M_{\rm p} = 1M_{\rm J}$ at several times after the Eddington time $t_{\rm edd} = 1,3,10,30\times t_{\rm edd}$.  We again assume a value $\alpha = 0.1$ for the disk viscosity, which results in $t_{\rm edd} \approx 80$ days (eq.~[\ref{eq:tedd}]).  Figure \ref{fig:disknuLnu} confirms the expectation from equation (\ref{eq:hnupeak}) that the spectrum initially peaks in the far UV at $t \gtrsim t_{\rm edd}$, but that $\nu_{\rm peak}(r = R_{\odot}) \propto \dot{M}_{\rm d}^{1/4} \propto t^{-1/3}$ decreases slowly thereafter (eq.~[\ref{eq:mdot}], [\ref{eq:hnupeak}]).

Figure \ref{fig:disklc} shows the light curves in various wavebands, ranging from far UV (FUV; $\nu = 2\times 10^{15}$ Hz) and near UV (NUV; $\nu = 1.4\times 10^{15}$ Hz) to K-band ($\nu = 1.4\times 10^{15}$ Hz) in the near infrared, shown now for the cases of a planet with mass $M_{\rm p} = 1M_{\rm J}$ and $M_{\rm p} = 10M_{\rm J}$ in the top and bottom panels, respectively.  Observe that at optical frequencies, the peak luminosity is $\nu L_{\nu} \gtrsim 10^{36}$ erg s$^{-1}$ on a timescale $\gtrsim t_{\rm edd}\sim $month$-$year, before fading as $\nu L_{\nu} \propto t^{-1}$ thereafter.  This is somewhat greater than the final luminosity from the super-Eddington phase, which we have added to the visual light curves at times $t < t_{\rm edd}$ using our results from Figure $\ref{fig:superEdd}$.  Comparing the predicted light curves at $t < t_{\rm edd}$ and $t > t_{\rm edd}$ suggests that the sub-Eddington transition is probably accompanied by an optical re-brightening (although the transition could be more subtle depending on the parameters).  At UV wavelengths the transition is clearly more dramatic, with the flux increasing from effectively zero at $t \ll t_{\rm edd}$ to $\nu L_{\nu} \sim 10^{37}$ erg s$^{-1}$ at $t \gtrsim t_{\rm edd}$.  

Finally, note that our model neglects the $\sim 1/2$ of the total gravitational energy released from the boundary layer connecting the disk and star.  Radiation from this layer will enhance the UV flux by a factor $\sim 2$ above our predictions in Figures \ref{fig:disknuLnu} and \ref{fig:disklc}.  

\section{Discussion and Conclusions}
\label{sec:discussion}

\subsection{Summary of Merger Transients}
\label{sec:summary}

Table \ref{table:summary} summarizes the predicted electromagnetic transients from planet-star mergers described in this paper.

In the case of direct-impact mergers ($\S\ref{sec:interior}$) the predicted transient is summarized by Figure \ref{fig:schematiclc} and quantified in Figure ~\ref{fig:radiation}.  Initially, emission is visible directly from the heated stellar surface and peaks at EUV/soft X-ray wavelengths.  At the earliest times ($\gtrsim$ months$-$year prior to merger) the signal is periodic on a timescale of hours, with a duty cycle $\sim 10$ per cent and a characteristic luminosity $\sim 10^{35}(\epsilon_{X}/0.1)$ erg s$^{-1}$ (with the exact light curve shape depending on the observing angle and geometry of the emitting region).  Periodicity ends at somewhat later times, but the orbit-averaged average luminosity continues to increase exponentially with time, before peaking at $L_{\rm EUV/X} \sim 10^{36}(\epsilon_{X}/0.1)$ erg s$^{-1}$ on a timescale $\sim$week$-$months prior to the merger.  

EUV/X-ray emission fades once outflows powered by the merger becomes sufficiently opaque to block radiation from the stellar surface (Fig.~\ref{fig:schematicwind}).  The final stages of the merger are instead characterized by optical emission from the outflow itself, which brightens from a luminosity $L_{\rm opt} \sim 10^{35}$ erg s$^{-1}$ on a timescale of a week$-$months prior to merger, to a peak luminosity $L_{\rm opt} \sim 10^{37}-10^{38}$ erg s$^{-1}$ on a timescale of days after merger (eqs.~[\ref{eq:Lpeak}], [\ref{eq:tpeak}]).  The predicted optical spectrum is that of a dense outflow with abundances characteristic of the stellar surface (metallicity $Z \sim Z_{\odot}$).  The velocity width is characteristic of the stellar escape speed $\sim$ hundreds km s$^{-1}$ and should remain relatively constant in time until after the optical peak (after which the line widths may decrease as the photosphere recedes to greater depths, and hence lower velocities, in the homologously expanding ejecta).

For tidal-disruption merger events ($\S\ref{sec:exterior}$, Figs.~\ref{fig:superEdd}, \ref{fig:disknuLnu}, \ref{fig:disklc}) emission initially peaks at optical wavelengths and originates from the super-Eddington outflow.  This emission brightens on a timescale $\sim$ days-week, before peaking at a luminosity $L_{\rm opt} \sim 10^{37}-10^{38}$ erg s$^{-1}$ and decaying as $\nu L_{\nu} \propto t^{-1}$ thereafter.  At UV wavelengths the emission is initially dim, but then brightens considerably to $\nu L_{\nu} \sim 10^{37}$ erg s$^{-1}$ on a timescale $t \sim t_{\rm edd} \sim$ months- few years, once the hot inner edge of the accretion disk becomes directly visible (Fig.~\ref{fig:disklc}).  Spectral line shapes should evolve from P-Cygni profiles characteristic of a dense outflow at times $t \ll t_{\rm edd}$, to emission lines characteristic of the hotter interior at $t \gtrsim t_{\rm edd}$.  The outflowing material likely has a super-solar metallicity ($Z\sim 5Z_{\odot}$), characteristic of the disrupted planet.  If the velocity of the outflow is proportional to the escape speed from the launching point of the wind, then one predicts that the line velocities will initially decrease as the accretion disk viscously expands, but then will increase again as the photon-trapping radius recedes towards the the stellar surface.  Once the disk itself becomes visible at times $t \gg t_{\rm edd}$, the emission lines may develop double-horned profiles characteristic of the Keplerian disk.

\subsection{Detection Prospects and Strategy}
\label{sec:detect}

A planet-star merger in our own galaxy, though intrinsically very bright, would be challenging to detect due to extinction through the Galactic plane.  A more promising detection strategy is via a systematic search of nearby galaxies at optical, UV, and X-ray wavelengths.  Since to first order the rate of planet-star mergers is proportional to the total number of stars, the most promising targets for such a search are the most massive nearby galaxies, such as M31 (Andromeda).  Indeed, since M31 and the Milky Way share a similar mass, one should expect a comparable merger rate, $\sim 0.1-1$ yr$^{-1}$ ($\S\ref{sec:freq}$).  A systematic survey of M31 over a timescale of years is thus one of the most promising ways to detect planet-star mergers. 

One important question is why planetary mergers have not yet been detected by existing surveys.  The XMM-Newton survey of M31 (\citealt{Pietsch+05}) identified 18 super-soft X-ray sources with luminosities $L_{\rm X} \gtrsim 10^{35}$ erg s$^{-1}$, none of which were obviously planet-star mergers.  However, since this survey covered $\approx 1.2$ square degrees (representing only a fraction of the stellar light of M31, which spans several degrees on the sky) and given the low predicted rate $\lesssim 0.1-1$ yr$^{-1}$ galaxy$^{-1}$ of direct-impact mergers, it is not particularly surprising that no candidate mergers were detected.  

A perhaps more promising instrument to search for direct-impact mergers is the upcoming satellite mission {\it eROSITA} \citep{Predehl+10}, which has a large field of view $\sim 1$ degree (allowing it to cover M31 in just a few pointings) and is scheduled to launch around 2013.  A one hour integrated exposure by {\it eROSITA} would reach a depth at soft X-rays $\sim 0.5-2$ keV of $\sim 10^{-14}$ erg cm$^{-2}$ s$^{-1}$, corresponding to a luminosity $L_{\rm X} \sim 10^{36}$ erg s$^{-1}$ at the distance of M31, comparable to the predicted peak brightness of direct-impact mergers given reasonably optimistic assumptions ($\epsilon_{\rm rad} \gtrsim 0.1$).  Although the all-sky survey performed by {\it eROSITA} will not reach this depth for $\approx 4$ years (much longer than the predicted peak duration of the X-ray emission $\sim$week$-$months from direct-impact mergers), a series of pointed observations over a shorter period could reach a substantially greater depth.  Candidate events could then be followed up with more sensitive observations, in order to track the light curve evolution or to search for periodicity.

Optical searches also provide a promising means for detecting planet-star mergers, in particular given the recent advent of sensitive synoptic surveys such as the Palomar Transient Factory (PTF; \citealt{Law+09}) and Pan-STARRs (\citealt{Kaiser+02}).  Advantages of an optical search include the fact that bright emission is predicted to accompany both direct-impact and tidal-disruption merger events, and that the predicted emission is more robust than in X-rays (the latter which depends sensitively on how efficiently radiation escapes the stellar surface).  The search for ``Fast Transients in Nearest Galaxies'' (P60-FasTING) survey performed with PTF \citep{Kasliwal+11} already surveys nearby galaxies with a survey cadence of $\sim 1$ day to magnitude limit $g \lesssim 21$ , corresponding to a luminosity depth $L_{\rm opt} \gtrsim 6\times 10^{36}$ erg s$^{-1}$ at the distance of M31, in principle sufficient to detect the merger of planets with masses $M_{\rm p} \gtrsim 1M_{\rm J}$. 

A major challenge of optical searches is the ability to distinguish planet-star mergers from unrelated astrophysical events which otherwise appear similar.  In particular, optical transients from both direct-impact and tidal-disruption mergers may in some ways resemble classical novae, which share similar durations and peak luminosities (e.g.~\citealt{Shafter+11}; \citealt{Kasliwal+11}) and are a factor $\gtrsim 50$ times more common than even our optimistic estimate of the planet-star merger rate.  One important difference is the characteristic outflow velocity, which is predicted to be $\sim 100-400$ km s$^{-1}$ in the case of planet-star mergers, but is typically much higher $\gtrsim 10^{3}$ km s$^{-1}$ in novae.  Of the $\sim 40$ novae compiled by \citet{Shafter+11} with measured H$\alpha$ line widths, only one (M31N 2007-11g) had a FWHM velocity $\lesssim 500$ km s$^{-1}$ (see their Fig.~16).  Observations of this event one year later, however, showed that it had re-brightened to approximately its original luminosity, indicating that it was probably a long-period Mira-like variable (\citealt{Shafter+08}, ATel 1851) instead of a nova or planet merger event.

Another distinguishing feature of novae is that they result from the ejection of a discrete shell of ejecta (powered by a thermonuclear runaway), followed by a relatively steady wind from the surface of the hot white dwarf.  For direct-impact mergers ($\S\ref{sec:interior}$), by contrast, we predict the opposite evolution$-$namely, optical emission that originates from a quasi steady-state wind at early times, before later transiting into a dynamical ejection just prior to the merger.  Tidal-disruption events ($\S\ref{sec:exterior}$) in some ways represent a closer analog to novae, in that the event evolves fastest initially, after which the `hot' central object (the inner accretion disk in the case of planet-star mergers, versus the hot white dwarf in novae) becomes increasingly visible with time.  The predicted light curve in this case (rise time $\sim$ days-week, followed by a subsequent decay $\nu L_{\nu} \propto t^{-1}$) also bears a qualitative resemblance to that of some novae (e.g.~\citealt{Kasliwal+11}; \citealt{Shafter+11}).  

Ultimately, the most promising strategy to detect and identify planet-star mergers is to combine optical, UV, and X-ray observations.  For instance, one way to distinguish novae from direct-impact planet mergers is to search for spatially coincident X-ray emission months$-$year prior to optical maximum.  Unfortunately, most searches to date have focused on X-ray emission following the nova (\citealt{Pietsch+05}; \citealt{Henze+10}), with only a few searches for `quiescent' emission prior to eruption (e.g.~\citealt{Orio+01}).  Likewise, a promising way to distinguish novae from planet tidal-disruption events is to search for spatially coincident bright UV emission months$-$years following optical maximum.  Multi-wavelength diagnostics, aided by systematic surveys of nearby galaxies, are clearly key to successfully identifying these relatively rare events.

\begin{table*}
\begin{center}
\vspace{0.05 in}\caption{Key Properties of Transients from Planet-Star Mergers}
\label{table:summary}
\resizebox{18cm}{!}
{
\begin{tabular}{cccccc}
\hline
\hline
\multicolumn{1}{c}{Event Type} &
\multicolumn{1}{c}{Photon Energy} &
\multicolumn{1}{c}{Peak Luminosity} &
\multicolumn{1}{c}{Peak Duration} &
\multicolumn{1}{c}{Ejecta Metallicity} &
\multicolumn{1}{c}{Characteristic Velocity} \\
 & & (erg s$^{-1}$) & & & \\
\hline
\hline
Direct-Impact ($\S\ref{sec:interior}$; Figs.~\ref{fig:schematic}, \ref{fig:radiation}, \ref{fig:schematiclc}): &  &  &  &  &  \\
\hline
Stellar Surface Emission ($\S\ref{sec:Xray}$) & EUV/Soft X-ray & $10^{36}(\epsilon_{\rm rad}/0.1)$ & months$-$year & N/A & N/A \\
Inspiral-Driven Outflow ($\S\ref{sec:optical}$) & Optical & $10^{37}-10^{38}$ (eq.~[\ref{eq:Lpeak}]) & days (eq.~[\ref{eq:tpeak}]) & $\sim Z_{\odot}$ & $\sim 100-400$ km s$^{-1}$ (blue shifted) \\
\hline 
\hline
Tidal-Disruption ($\S\ref{sec:exterior}$; Figs.~\ref{fig:superEdd}, \ref{fig:disknuLnu}, \ref{fig:disklc}): &  &  &  &  &  \\
\hline
Super-Eddington Disk Wind ($\S\ref{sec:superEdd}$) & Optical & $10^{36}-10^{37}(M_{\rm p}/M_{\rm J})$ & day$-$week & $\sim 5Z_{\odot}$ & $\sim 100-400$ km s$^{-1}$ (blue shifted) \\
Accretion Disk ($\S\ref{sec:subEdd}$) & Optical(UV) & $10^{36}(10^{37})$ & months$-$year & N/A & $\sim 100-400$ km s$^{-1}$ (double horned) \\
\hline
\hline
Classical Novae$^{(a)}$ & Optical & $10^{37}-10^{39}$ & days$-$months & $\sim 3-30Z_{\odot}^{(b)}$ & $\gtrsim 10^{3}$ km s$^{-1}$ (blue shifted) \\
\hline
\hline
\end{tabular}
}
\end{center}
{$^{(a)}$Shown for purposes of comparison; $^{(b)}$\citet{Livio&Truran94}.}
\end{table*}
 
\subsection{Conclusions and Future Work}
\label{sec:conclusions}

Basic considerations suggest that planets merge with their host stars at a relatively regular rate ($\S\ref{sec:freq}$; Fig.~\ref{fig:cumsum}).  Detecting such events directly, beyond confirming the basic physical picture outlined here, could provide key information on the properties of the merging binary, as well as inform the rate of tidal dissipation (i.e.~the value of $Q'_{\star}$) and/or the rate at which planets currently flow into their inner stellar system (\citealt{Socrates+11}).  

In $\S\ref{sec:outcomes}$ we identified three qualitatively different merger outcomes (`direct-impact', 'tidal-disruption' and 'stable mass transfer'), which, from the current sample of explanets, we estimate that an appreciable fraction of systems will experience each (Figs.~\ref{fig:demographics}; Table \ref{table:outcomes}).  Our estimate of the relative fraction of merger outcomes assumed that the current masses and radii of extrasolar planets and their host stars represent their values at the point of merger, an assumption of questionable validity.  Although in $\S\ref{sec:outcomes}$ we described the effects of late stellar evolution on the frequency and outcome of planet-star mergers, the properties of the planet itself may change even when the star is still on the main sequence.  As a planet migrates inwards towards merger, it experiences increased interior heating (e.g.~tidal dissipation, ohmic heating, atmospheric irradiation) which may inflate its radius over the nominal value given its age (e.g.~\citealt{Bodenheimer+01}; \citealt{Burrows+08}; \citealt{Ibgui&Burrows09}; \citealt{Batygin+11}).  Close-in planets may furthermore lose mass to outflows powered by stellar irradiation, although current models suggest that hot Jupiters are unlikely to completely evaporate by this process (e.g.~\citealt{Murray-Clay+09}, \citealt{Adams11}; though less massive planets may be more susceptible to evaporation, e.g.~\citealt{Rappaport+12}).  Given these uncertainties, it is difficult to determine with confidence which merger outcome is the most frequent.

In this paper we have focused primarily on the merger of planets in quasi-circular orbits, since most known hot Jupiters should merge with final eccentricities $e \lesssim 10^{-3}$ (Appendix $\ref{app:B}$).  Such a small eccentricity would not qualitatively change our conclusions regarding the dynamics of inspiral described in $\S\ref{sec:dynamics}$ since the difference between pericenter and apocenter is smaller than the stellar scale height $H \gtrsim 10^{-3}R_{\star}$.  Nevertheless, depending on the rate at which planets are deposited directly into the inner stellar system by planet scattering or Kozai oscillations (or to what extent gravitational interactions with other planets pump eccentricity), some planets may enter with low impact parameters on highly eccentric orbits.  

Depending on the impact parameter of the trajectory and the mass ratio of the system, an eccentric planet-star merger can result in either a direct head-on collision; complete ejection of the planet; or a tidal-disruption event (\citealt{Guillochon+11}).  In the case of a direct head-on collision, the transient may resemble a shortened version (or final stages) of a direct-impact merger ($\S\ref{sec:interior}$), possibly producing a prompt flash of high-energy emission from the stellar surface (see also \citealt{Zhang&Sigurdsson03}) or thermal emission from an impact-driven outflow ($\S\ref{sec:optical}$).  In the tidal-disruption case, the emission would qualitatively resemble the quasi-circular merger case that we have already discussed ($\S\ref{sec:exterior}$), except that the late-time accretion rate (and hence the resulting optical-EUV light curve) will decrease differently with time $\dot{M} \propto t^{-5/3}$ due to fall-back accretion \citep{Rees88}, than the evolution $\dot{M} \propto t^{-4/3}$ predicted due to the viscous evolution of the disk (eq.~[\ref{eq:mdot}]).

In addition to classical novae, transient emission from planet-star mergers may in some ways resemble that from the merger between two binary stars, events which may occur at a similar rate.  The optical transients M85 OT2006-1 and V838 Mon have both been argued to result from stellar mergers (\citealt{Munari+02}; \citealt{Soker&Tylenda03,Soker&Tylenda06}; \citealt{Kulkarni+07}), although other explanations have been proposed for these events (e.g.~\citealt{Retter&Marom03}; \citealt{Pastorello+07}).  The recent OGLE transient V1309 Scorpii represents an essentially indisputable case of a binary stellar merger \citep{Tylenda+11}.

One might think that a merger between two stars would be substantially brighter than a planet-star merger due to the considerably greater gravitational energy released during such an event.  However, this expectation may not necessarily be realized.  One ``advantage'' enjoyed by low mass-ratio mergers is that insufficient energy is released during the earlies stages of the merger to appreciably expand the outer stellar envelope.  In this case the heated atmosphere of the star may have sufficient time to cool before the next planetary orbit, thus increasing the likelihood that high energy emission may escape the viscnity of the stellar surface ($\S\ref{sec:Xray}$), at least during the earliest stages of the merger.

A perhaps more important advantage of planetary versus stellar mergers is that, in binaries with mass ratios $q \gtrsim 0.1$ the secondary is sufficiently massive to maintain synchronous rotation of the stellar envelope of the primary until the point of merger (\citealt{Soker&Tylenda06}).  This implies that substantially less relative kinetic energy is available between the merging stellar surfaces to power prompt emission than in the case of a less massive secondary (\citealt{Soker&Tylenda06}).  The quantity of mass ejected, and the resulting prompt electromagnetic display, could thus in principle be much less luminous than one would anticipate based on a simple `scaled-up' version of the model presented in $\S\ref{sec:radiation}$ (in which we have implicitly assumed a slowly rotating star).  

Despite these caveats, to the extent that stellar mergers eject {\it some} quantity of ionized material, much of the same physics described in $\S\ref{sec:interior}$ and $\S\ref{sec:exterior}$ regarding optical transient emission should apply to these events as well.  For instance, V1309 reached a peak optical luminosity $\sim 10^{38}$ erg s$^{-1}$ lasting for a timescale $\sim 1$ month, such that the total radiated energy was $\sim 3\times 10^{44}$ erg.  If this luminosity were powered primarily by the recombination of hydrogen in an outflow (as we find characterizes our outflow models in $\S\ref{sec:optical}$ and $\S\ref{sec:superEdd}$), then the requisite mass of ejecta is $\sim 0.01M_{\odot}$, a quantity in agreement with independent estimates of similar events (\citealt{Tylenda&Soker06}).  

Much additional work is required to better understand the observable signatures of planet-star mergers.  In the case of direct-impact mergers, a more detailed model of the hot post-planetary layer ($\S\ref{sec:hotlayer}$) is necessary to make more quantitative predictions, the study of which will likely require a detailed radiation hydrodynamical calculation.  Our basic estimates hint that a sizable fraction of the thermal energy generated near the surface may escape as high energy radiation; however, a better understanding of the detailed structure of this region (including the effects of sustained heating in the vicinity of the planet-star interface) is necessary to assess the radiative efficiency $\epsilon_{\rm rad}$, geometry, and beaming fraction of the resulting emission (Fig.~\ref{fig:schematic}).  Another major uncertainty is the extent to which the heated stellar surface is able to cool, or to what extent the star remains `puffed up', by the time the planet has completely orbited the star.  Our calculations in $\S\ref{sec:dynamics}$ assumed that the planet interacts with an essentially unperturbed stellar atmosphere after each passage, clearly an oversimplification, especially during the final dynamical stages of the merger.  Depending on the detailed structure of the perturbed atmosphere, this could accelerate (or decelerate) over the inspiral evolution calculated in $\S\ref{sec:dynamics}$.  

Given the sensitive, wide-field capabilities of surveys at X-ray and optical wavelengths, it seems promising that planet-star mergers should be detected within the next decade.

\section*{Acknowledgments}
BDM is supported by NASA through Einstein Postdoctoral Fellowship grant number PF9-00065 awarded by the Chandra X-ray Center, which is operated by the Smithsonian Astrophysical Observatory for NASA under contract NAS8-03060.  DG acknowledges support from from the Fermi 4 Cycle grant number 041305.  DSS gratefully acknowledges support from NSF grant AST-0807444 and the Keck Fellowship. 

\begin{appendix}

\section{Stable Vs. Unstable Mass Transfer}
\label{app:A}

Here we quantify the fate of a planet that fills its Roche-lobe outside the stellar surface, i.e., for which $a_t>R_{\star}+X_t$ (eq.~[\ref{eq:cond1}]), where $a_t$ (eq.~[\ref{eq:at}]) is the Roche radius and $X_t$ is the distance of the L1 Lagrange point from the center of mass as described in $\S\ref{sec:outcomes}$.  Our goal is to explore whether infinitesimal mass loss from the planet $dM_{\rm p}<0$ through the Lagrange L1 point results in the planet expanding faster or slower than the size of Roche-lobe, in order to determine whether mass transfer is stable or unstable.\footnote{For the extreme mass ratio systems under consideration, the distance from the center of the planet to the L2 Lagrange point is similar to that to L1 (e.g.~\citealt{Gu+03}; \citealt{Nayakshin&Lodato11}).  Some mass will therefore overflow from L2 as well.  Mass loss from L1 is, however, still likely to dominate since the star is strongly irradiated from the front, resulting in a larger atmospheric scaleheight near L1.  Since mass loss from L2 results in angular momentum loss from the system, its effect is to increase the likelihood of unstable mass transfer and dynamical tidal-disruption (see below).}

Equating the specific angular momentum $j_{\rm ov}=(GM_{\star}/a_t)^{1/2}(a_t-X_t)$ of matter leaving the planet with that of a Keplerian orbit about the star $= (GM_{\star}R_c)^{1/2}$ we obtain the characteristic circularization radius of the material
\be
R_{\rm c}=a_{\rm t}(1-X_t/a_t)^2.
\ee

If $R_{\rm c}<R_{\star}$ then the accretion stream directly impacts the stellar surface and no disk forms, such that all of the mass and angular momentum carried by $dM_{\rm p}$ is transferred directly to the star.  In the much more common case that a disk forms (i.e.~$R_{\rm c} > R_{\star}$) it is still reasonable to assume that most of the transferred mass will ultimately be accreted the star;\footnote{Although this need not always be the case, especially in the case of eccentric disruptions (\citealt{Guillochon+11}; \citealt{Nayakshin&Lodato11}).} however, the fate of the angular momentum in this case is less clear.

Assuming that the disk is Keplerian, its specific angular momentum near the stellar surface is $j_{\rm im}=(GM_{\star}R_{\star})^{1/2}$.  Thus only a fraction $f=j_{\rm im}/j_{\rm ov}=(R_{\star}/R_c)^{1/2}$ of the angular momentum of the accreting material is transferred onto the star.  The rest remains in the disk, where it will eventually be returned to the orbit via tides exerted on the planet.

Mass loss reduces the angular momentum of the planet by an amount $dJ_{\rm p}=dM_{\rm p}(GM_{\star}a)^{1/2}+M_{\rm p}(GM_{\star}/a)^{1/2}da/2$.  Equating this to the angular momentum gained by the star $dJ_{\rm p}=dJ_{im}=dM_{\rm p}(GM_{\star}R_{\star})^{1/2}$ results in the following expression
\be
\frac{{\rm d}a}{{\rm d}M_{\rm p}}=-2\frac{a}{M_{\rm p}}\big[1-\big(\frac{R_{\star}}{a}\big)^{1/2}\big].
\ee
The volume of the Roche-lobe scales as $V_{RL}\propto M_{\rm p}a^3$ or
\be
\frac{{\rm d}V_{\rm RL}}{{\rm d}M_{\rm p}}=\frac{V_{\rm RL}}{M_{\rm p}}\big(1+3\frac{M_{\rm p}{\rm d}a}{a{\rm d}M_{\rm p}}\big)=
\frac{V_{\rm RL}}{M_{\rm p}}\left[1-6\big(1-\big(\frac{R_{\star}}{a}\big)^{1/2}\big)\right].
\label{eq:dVRL}
\ee
For $a=a_t$ (eq.~[\ref{eq:at}]) the bracketed quantity in the last expression takes the value $[3\sqrt{2}(\bar{\rho}_{\rm p}/\bar{\rho}_{\star})^{1/6}-5]$.  Equation (\ref{eq:dVRL}) shows that if overflow occurs at large orbital separations ($a_t>>R_{\star}$ or $\bar{\rho}_{\rm p}/\bar{\rho}_{\star}<<1$), then the volume of the Roche-lobe increases rapidly with mass loss, while for $a_t\sim R_{\star}$ its size evolves more gradually.    

The change in the radius of a planet that undergoes mass loss depends on its entropy and mass, i.e.~$R_{\rm p}(M_{\rm p},S)$.  The planetary models of \citet{Spiegel&Burrows12} suggest that the relatively large observed radii of close-in planets indicate high entropies $S\sim 9k_B/$baryon.  If one assumes that the planet loses mass at constant entropy, then the analytic fits to the mass-radius relation given by \citet{Spiegel&Burrows12} give the following expression:
\be
\frac{{\rm d}V_{\rm p}}{{\rm d}M_{\rm p}}=C \frac{V_{\rm p}}{M_{\rm p}},
\ee
where $C=3\ln 10[(p_{01}+p_{11}S+p_{21}S^2)(M_{\rm p}/M_{\rm J})+2(p_{02}+p_{12}S) (M_{\rm p}/M_{\rm J})^2]$; $S$ is in units of $k_B/$baryon; and the parameters $p_{ij}$ are given in  \citet{Spiegel&Burrows12}.  Since typical values are $C\sim -1$ for $M_{\rm p}\sim 1 M_{\rm J}$ and $C\sim -0.3$ for $M_{\rm p}\sim 10 M_{\rm J}$, we conclude that most of the observed planets expand upon mass loss.  

Since the masses of the planets in our sample are measured, we can directly estimate the value of $\frac{{\rm d}V_{\rm p}}{{\rm d}M_{\rm p}}$ for individual systems that will undergo Roche-lobe in order to predict their ultimate fate.  In particular, if $\big|\frac{{\rm d}V_{\rm p}}{{\rm d}M_{\rm p}}\big|>\big|\frac{{\rm d}V_{\rm RL}}{{\rm d}M_{\rm p}}\big|$ then the planet will expand faster than the Roche-lobe with mass loss, indicating that mass transfer is unstable (and vice versus).  Figure \ref{fig:demographics} summarizes our results applying this analysis to infer the stability of mass transfer for the entire sample of transiting planets (assuming that the properties of the planet and star today reflect those at the time of merger).  

\section{Justification of Circular Orbits for Direct-Impact Mergers}
\label{app:B}

Here we analyze the tidal evolution of star-planet systems in order to derive the orbital decay timescale and to estimate the eccentricity of systems just prior to merger or tidal-disruption.  We focus on a compact planet-star systems and ignore the effects of potential interactions with other planets (e.g.~eccentricity `pumping'). 

Assuming that the planet resides in a synchronous orbit about a relatively slowly rotating star, the evolution of the orbital parameters of the system $e$, $a$ are given by the expressions (in the low eccentricity $e\simless 0.2$ limit; see \citealt{Goldreich&Soter66}; \citealt{Kaula68}; \citealt{Peale&Cassen78}; \citealt{Jackson+08}; \citealt{Barnes+09}; \citealt{Ibgui+10})   
\be
\frac{1}{e}\frac{{\rm d} e}{{\rm d} t}=-a^{-13/2}[K_1R_{\rm p}^5/Q'_{\rm p}+K_2R_{\star}^5/Q'_{\star}],
\ee
\be
\frac{1}{a}\frac{{\rm d} a}{{\rm d} t}=-a^{-13/2}[2K_1R_{\rm p}^5e^2/Q'_{\rm p}+8K_2R_{\star}^5/25Q'_{\star}],
\ee
where $K_1=(63/4)G^{1/2}M_{\star}^{3/2}/M_{\rm p}$, $K_2=(225/16)G^{1/2}M_{\rm p}/M_{\star}^{1/2}$, and $Q'_{\rm p}$ is the modified tidal quality factor of the planet; the rest of the quantities are defined in Section $\S\ref{sec:outcomes}$. For a discussion of the several underlying assumptions in deriving these expressions, see \citet{Ibgui+10}. 

Tides raised on the planet (terms containing $Q'_{\rm p}$) serve to circularize the orbit. 
For sufficiently low $Q'_{\rm p}$, the orbit circularizes on a timescale shorter than that of orbital decay. 
The fact that the shorter period planets with $P\simless 3$ days 
are in near circular orbits ($e\simless 0.1$, in contrast to wider systems that can be very eccentric) indicates that circularizarion tides operate effectively in the tight star-planet systems (e.g.~\citealt{Jackson+08}).  If this is the case, then both tidal-disruptions and mergers take place when the orbit is nearly circular.

An {\it upper limit} on the final eccentricity of the orbit is derived by setting $Q'_{\rm p}\to \infty$, in which limit $a$ and $e$ decay on the same timescale. The ratio of expressions B1/B2 in this limit leads to d$e/e=(25/8){\rm d} a/a$. Integrating the last expression starting from the current values $a_{\rm in}$, $e_{\rm in}$ one finds $e_{\rm f}=e_{\rm in}(a_{\rm f}/a_{\rm in})^{25/8}$.  The most compact systems with $a_{\rm in}\sim (6-8)R_\odot$ have observed $e_{\rm in}\simless 0.1$.
Setting $a_{\rm f}=R_\odot$ (merger case) gives $e_{\rm f}\simless 3\times 10^{-4}$ while for $a_{\rm f}\simeq 2R_{\odot}$
the final eccentricity of the orbit is $e_{\rm f}\simless 3\times 10^{-3}$. 
These small values of eccentricity at the point of tidal-disruption or merger are not expected to appreciably change the analysis presented in $\S\ref{sec:interior}$, which implicitly assumed perfectly circular orbits.  This is because the difference between pericenter and apocenter $\sim 2e_{\rm f} \lesssim 10^{-3}$ is smaller than the stellar scale height $H \gtrsim 10^{-3}R_{\star}$.

Exression B2 also provides an estimate for the orbit tidal decay timescale (again setting  $Q'_{\rm p}\to \infty$) as given in equation (\ref{eq:plungetime})

\end{appendix}

\bibliographystyle{mn2e}
\bibliography{ms}


\end{document}